\documentclass[sigconf,nonacm]{acmart}
\usepackage{tabularx}
\usepackage{tikz}
\usepackage{pgfplots}
\usepackage{subcaption}
\pgfplotsset{compat=1.18}
\AtBeginDocument{%
  }

\setcopyright{none}
\renewcommand\footnotetextcopyrightpermission[1]{}

\begin{document}

\title  {Encypher: Shared Agency and Social Presence in Collaborative Music Generation for Dance Cyphers}

\author{Zhixing Chen}
\email{zhixingc@mit.edu}
\orcid{0009-0009-7222-7582}
\affiliation{%
  \institution{Massachusetts Institute of Technology}
  \city{Cambridge}
  \country{USA}
}

\author{Cheng-Zhi Anna Huang}
\email{huangcza@mit.edu}
\orcid{0000-0002-8440-2378}
\affiliation{%
  \institution{Massachusetts Institute of Technology}
  \city{Cambridge}
  \country{USA}
}

\renewcommand{\shortauthors}{Chen and Huang}

\begin{abstract}
  Music and dance are social practices of expression and connection, yet most HCI work in human-AI co-creation centers the solo performer. As generative music matures, we ask not only what AI can compose but what social encounters it can organize around sound. We present Encypher, a collaborative generative music system that translates collective movement qualities into text prompts conditioning real-time music generation for dance cyphers. Through five weeks of co-design with local dancers, a user study with unacquainted participants, a public museum event, and a live performance, we found that users developed shared agency, perceiving the music as a response to the room's energy. While newcomers felt uncertain, the system fostered social presence by prompting them to look to each other for cues. By treating sociality as a design concern rather than a downstream effect, we offer a framework and design implications for AI systems for collaborative, embodied expression.
\end{abstract}


\begin{teaserfigure}
  \includegraphics[width=\textwidth]{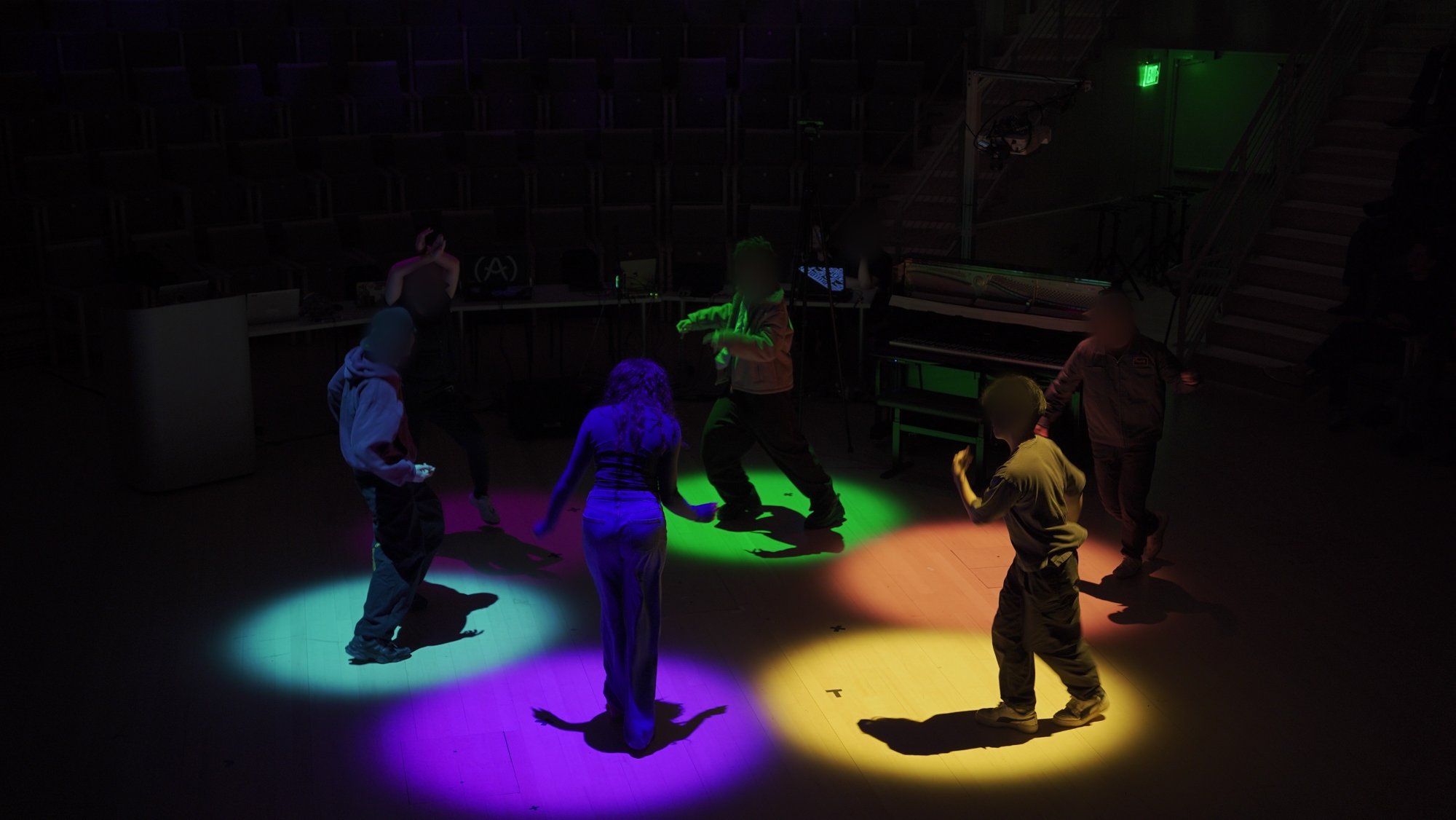}
  \caption{Dancers in an Encypher cypher, each standing in a colored pool of light projected onto the floor of their zone; the screen has been replaced by light on the dancers themselves.}
  \Description{Six people dancing in a dark performance hall, each lit from above by a differently colored spotlight forming a circle of colored pools on the floor.}
  \label{fig:teaser}
\end{teaserfigure}
\maketitle

\section{Introduction}

Music and dance are ways people coordinate attention, express identity, and meet one another through the body. From early instruments to turntables and contemporary sampling devices, technology has continually reshaped how we create, perceive, and share rhythm and movement. Artificial intelligence (AI) extends this lineage: once limited to narrowly structured tasks, AI is now a creative medium capable of perceiving, generating, and co-creating with humans across modalities. Multimodal systems can now capture the gestures of the body and the intricacies of music together, opening new ways of communicating through movement \cite{chang_be_2025}.

Yet as generative music becomes more musically usable, most HCI work in this space still centers the solo performer. We see an opportunity to ask not only what AI can compose, but what kinds of social situations it helps create---facilitating human-to-human connection rather than only human-AI interaction.

\begin{figure}
    \centering
    \includegraphics[width=1\linewidth]{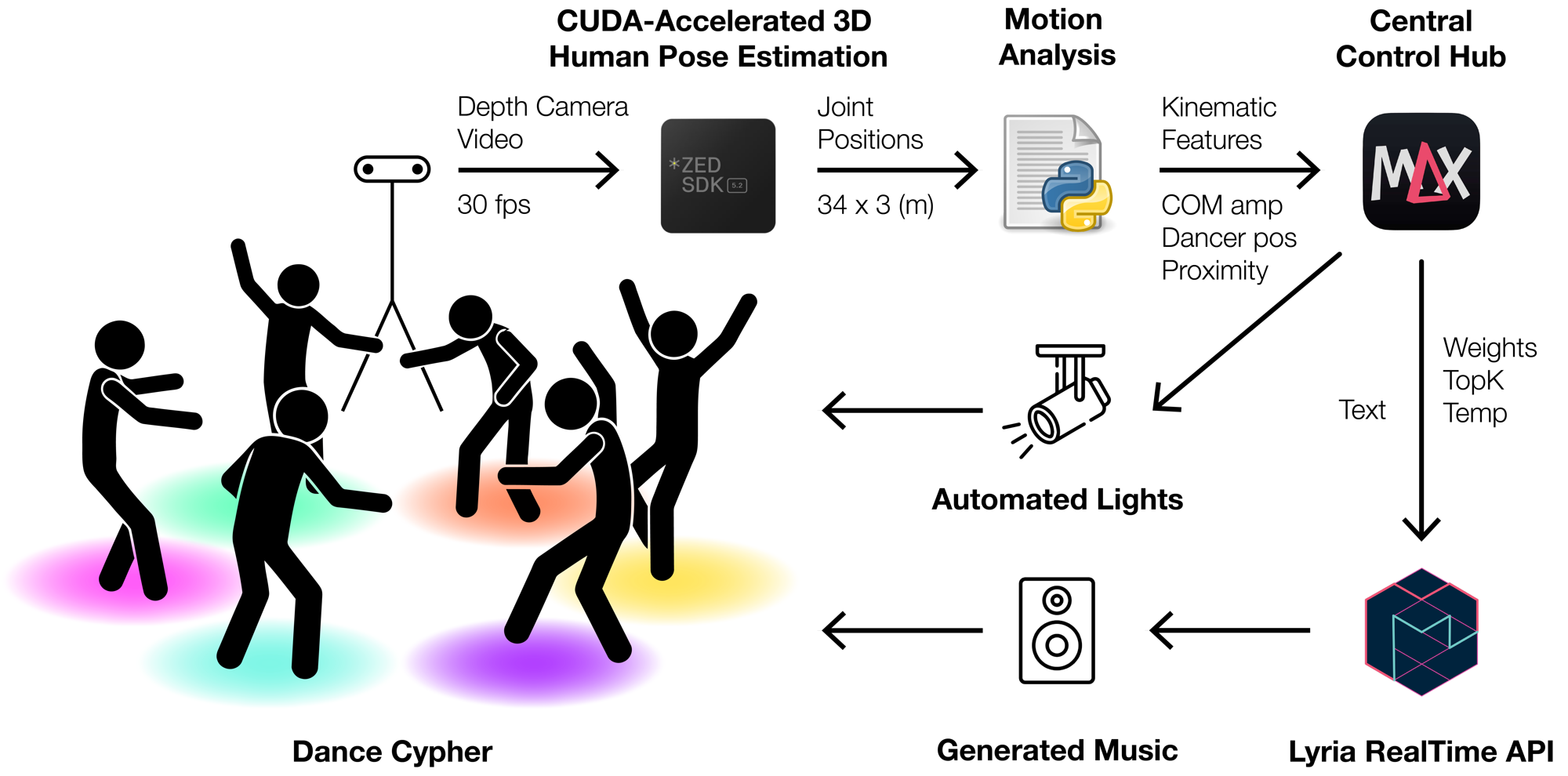}
    \caption{The Encypher system uses a depth camera to capture the joint movements of dancers in a cypher. Motion analysis extracts kinematic features such as center of mass amplitude, dancer position, and proximity, and sends them to Max/MSP. Max/MSP serves as a central control hub for music generation and lighting. Kinematic features control the weights, top-K, and temperature of musical elements in the LyriaRT API to generate music in real time; center-of-mass energy also controls light intensity.}
    \Description{System diagram. A depth camera films dancers standing in a circle; motion analysis extracts kinematic features and sends them to Max/MSP, which controls prompt weights, top-K, and temperature in the LyriaRT API and drives stage lighting; the generated music plays back to the dancers.}
    \label{fig:hero_diagram}
\end{figure}

We present Encypher\footnote{\url{https://encypher-chi.github.io/}}, a collaborative generative music system that translates collective movement qualities into text prompts to condition real-time music generation for dance cyphers (Figure~\ref{fig:hero_diagram}). In Hip-Hop culture, a cypher is a circular freestyle practice defined by energy sharing, exchange, and communal response. The name Encypher plays on \textit{encipher}, reflecting the system's role in translating embodied movement into a generative musical process. Encypher builds on \textit{Con Moto} \cite{chen_con_2026}, which used inference-time steering of a symbolic music transformer for dyadic dance improvisation. Where \textit{Con Moto} operated on note-level MIDI manipulation between two dancers, Encypher pivots to high-fidelity neural audio and expands the interaction into a true cypher, where all participants collectively steer music generation in real time.

\begin{figure}
    \centering
    \includegraphics[width=0.55\linewidth]{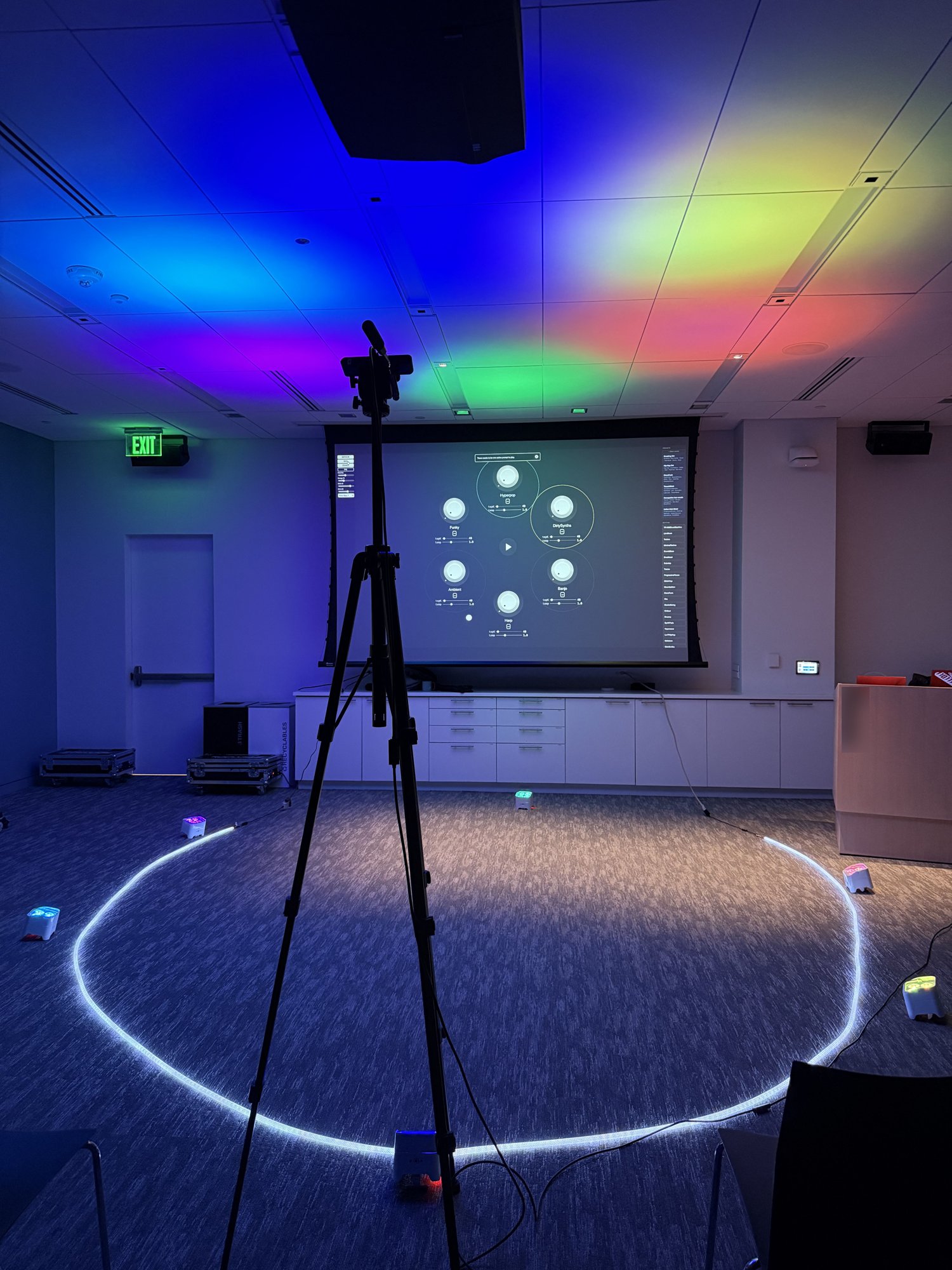}
    \caption{Encypher setup at a public museum event open to the broader community, with colored lights projected on the ceiling to reflect the zones, a depth camera to capture movement, and a visualization on screen.}
    \Description{Wide shot of a dark room at a museum event: a ring of colored lights on the ceiling, a depth camera on a tripod, and a large screen showing six circular zones.}
    \label{fig:museum_setup}
\end{figure}

We developed Encypher through autobiographical design based on the first author's own dance practice, five weeks of co-design with local dancers (n=15), a user study with unacquainted participants who are new to dance (n=11), a public museum event open to the broader community (Figure~\ref{fig:museum_setup}), and a screen-free public live performance in which light on the dancers replaced the visualization (Figure~\ref{fig:teaser}). Our findings reveal a shift from one-to-one control to shared agency: users perceived the music as responding to the room's energy rather than to their own movements alone. While novices and strangers often felt uncertain at first, the system's mappings brought them together to dance, bond, and learn from one another, supporting social presence by prompting them to look to each other for cues. The system acted as a social bridge for strangers in particular, though established dance communities such as breakers---with their own existing social dynamics---sometimes resisted its framing.

Our key contributions are:
\begin{itemize}
    \item \textbf{A collaborative generative music system, Encypher,} that maps collective movement qualities to text prompts conditioning real-time neural audio generation, designed to support the social dynamics of dance cyphers rather than solo performance.
    \item \textbf{Empirical findings on shared agency} from co-design sessions, a controlled user study, and public deployments, showing how generative music can shift dancers' perception of control from individual to collective and act as a social bridge among unacquainted participants.
    \item \textbf{Design implications for collaborative human-AI creative systems,} including the limits of imposed mappings on established communities, and a framework for foregrounding sociality as a central design concern in multi-modal human-AI interaction.
\end{itemize}

\section{Related Work}
Encypher sits at the intersection of three areas: the social dimensions of music and dance, technology-mediated collaboration, and generative systems for movement-driven music.

\subsection{Social Interaction}

\paragraph{Through Music}

Humans are social creatures, and music has long organized collective life. Archaeological records date musical instruments back 42,000--43,000 years, and contemporary electronic dance music concerts show how strongly rhythm still gathers bodies in shared space \cite{munzel_geisenklosterle_2002, volpe_measuring_2016, redfield_electronic_2017}. Music functions not only as entertainment but also as a way to build social bonds, negotiate group identity, and facilitate cooperation within and between communities \cite{hunt_sociology_2017}. Early musical activities likely paralleled the emergence of language, serving roles in social cohesion, ritual, and collective memory, while also supporting communication in contexts where spoken language was insufficient or unavailable.

\paragraph{Through Dance}

Dance, like music, organizes social life through the body. From an anthropological perspective, dance functions not merely as movement but as a cultural practice and social ritual that establishes ties and creates specific structures within communities \cite{pusnik_introduction_2010}. In traditional societies, dance marks life transitions, strengthens kinship ties, and maintains cultural continuity across generations \cite{penniman_rhythm_2002}. Beyond these cultural functions, dance also has interpersonal bonding effects. Modern research demonstrates that synchronized dancing significantly enhances social bonding between participants, which appears to activate endorphin systems in the brain, resulting in measurable increases in pain thresholds and positive social feelings toward dance partners \cite{tarr_synchrony_2015, weinstein_enhancing_2016}.

In particular, Hip-Hop is a style of music, dance, and culture that emerged in the Bronx, New York City in the 1970s, and has since spread across the world to become a global cultural phenomenon \cite{benvenga_hip-hop_2022}. Built on the idea of exchange, Hip-Hop centers the sharing of identity, skill, and available resources. When MCs rap, they tell stories and convey messages, a powerful form of self-expression. When dancers enter the cypher, a group circle where people take turns showing their moves, they share identity and technique with others.

\paragraph{Framework for Social Interaction}
Understanding how people coordinate, evaluate, and engage in social interactions requires frameworks that span psychological, sociological, and technological dimensions. At the most fundamental level, research in social perception characterizes interactions along two primary dimensions: warmth (trustworthiness, friendliness, and intention) and competence (capability, assertiveness, and skill) \cite{cuddy_warmth_2008, simon_social_2020}. In performance, observers rapidly judge both skill (competence) and qualities like generosity and authenticity (warmth), shaping cooperation and engagement. Rapid social evaluations contribute to the experience of social presence---the perception that others are real, attentive, and engaged in the interaction \cite{kreijns_social_2022}. Social presence reflects the felt experience of warmth and competence in action. Effective interaction balances capability with positive intent, a dynamic that technologies aiming to support social interaction must consider.

Although much HCI research focuses on remote communication, we focus on collocated interaction, where people share physical space. \citet{olsson_technologies_2020} identified three functional roles that technology can play: facilitating, inviting, and encouraging. Technology can make interaction easier, signal the possibility of socializing, and incentivize starting and maintaining the interaction. These roles extend beyond simply enabling communication to actively improving the quality or extent of social engagement. \citet{hornbaek_what_2019} distinguish between interaction style (the modality, structure, and means of interaction, such as embodied, multimodal, or spatial) and interaction quality (the subjective experience of naturalness, effectiveness, and expressiveness). Separating style from quality makes it possible to understand how specific design features influence people's experience and what makes interactions successful.

Group size also matters. \citet{simmel_sociology_1950} showed that dyads, triads, and larger ensembles are qualitatively different social forms: dyads are intimate but fragile, triads admit coalitions, and larger groups depend on alignment of skill, effort, and mutual awareness to coordinate. A cypher of five or six sits in this last regime.

A synthesis of these frameworks yields the integrated evaluation structure for understanding social interaction in music and dance shown in Table~\ref{tab:social_interaction_framework}. The dimensions span from core social-psychological processes to specific interaction qualities and the technological factors that support them.
\begin{table*}[t]
\centering
\caption{Dimensions for evaluating social interaction}
\label{tab:social_interaction_framework}
\begin{tabularx}{\textwidth}{p{3cm}Xp{4cm}p{4cm}}
\toprule
\textbf{Evaluation Dimension} & \textbf{Definition} & \textbf{Example Indicators} & \textbf{Relevance to Performance Contexts} \\
\midrule
Social Presence & Warmth, competence, feeling that others are present \cite{kreijns_social_2022, cuddy_warmth_2008, olsson_technologies_2020}& Eye contact, mutual gaze, attention  & Keeping track of all cypher participants \\
\addlinespace
Group Cohesion & Commitment to group goals, mutual attentiveness \cite{simmel_sociology_1950, olsson_technologies_2020}& Cooperation, synchrony, responsiveness & Maintaining cypher energy and inclusiveness \\
\addlinespace
Style & How interaction is structured and performed \cite{hornbaek_what_2019}& Gestures, movement, spatial patterns & Using embodied movement to communicate \\
\addlinespace
Quality & How the interaction feels \cite{hornbaek_what_2019} & Ease, flow, expressiveness & Energy or vibe in performance \\
\addlinespace
Facilitation & How environment or tech makes participation easier \cite{olsson_technologies_2020} & Accessibility, clear norms, low barriers & Circular configuration inviting entry \\
\addlinespace
Invitation & How cues encourage participation \cite{olsson_technologies_2020}& Turn cues, rhythmic breaks, gestures & One dancer signaling another's turn \\
\addlinespace
Encouragement & How participation is sustained \cite{olsson_technologies_2020}& Feedback, audience response, reinforcement & Crowd cheers motivating continued engagement \\
\bottomrule
\end{tabularx}
\end{table*}

\subsection{Technology-Facilitated Collaboration}

Music games have made co-performance accessible through consumer devices. Guitar Hero and Rock Band, for example, let multiple players share a virtual stage through simplified guitar, bass, drum, and vocal controllers in real time \cite{harmonix__redoctane_guitar_2005, harmonix__mtv_games_rock_2007}. In public art, the reacTable is a round tangible tabletop synthesizer designed for collaborative use \cite{jorda_reactable_2005}. Its creators note that even novices can pick it up quickly, and the circular table allows many people to gather and make music together.

Commercial dance games offer a parallel history for movement. Dance Dance Revolution, the arcade rhythm game that pioneered interactive dance gaming, creates physical spaces where players perform synchronized footwork patterns on pressure-sensitive pads \cite{konami_dance_1998}. Later franchises like Just Dance and Dance Central use motion sensors and cameras to allow multiple players to perform choreographed routines alongside on-screen dancers \cite{ubisoft_just_2009, harmonix__mtv_games_dance_2010}. \citet{gao_impact_2012} show that synchronized dancing in dance games facilitates meaningful social bonding by activating endorphin systems.

Social media has extended collaborative music and dance beyond physical spaces. TikTok's Duet feature chains asynchronous performances from multiple creators, supporting what \citet{otoole_collaborative_2023} calls ``distributed creativity.'' During the COVID-19 lockdown, communities such as JazzTok used Duet chains to simulate virtual jazz orchestras \cite{kaye_jazztok_2023, kaye_please_2022}.

These systems show how technology can bring people into shared rhythm, but their interaction is usually bounded: players follow charts, routines, platform formats, or fixed targets. For Encypher, the relevant question is what remains when there is no score to maximize and no choreography to copy. Collaborative creativity requires enough structure for people to enter together, but enough openness for them to negotiate musical and bodily choices in the moment. This is where real-time generative music becomes useful: it can give a group something shared to respond to while still allowing the sound to change with their collective behavior.

\subsection{Movement Sonification}
Sonification is traditionally defined as the functional transformation of data and interactions into sound \cite{hermann_sonification_2011}. In dance, the transformation typically involves translating kinetic data---such as joint orientation, velocity, or acceleration---into musical parameters \cite{hattwick_forming_2014, bang_suspended_2023, lyu_between_2025, kikuchi_transcriptions_2025}. Current approaches move beyond these structural mappings to treat sonification as a medium for what
\citet{bevilacqua_sensori-motor_2016} describe as sensori-motor enhancement. We ground our research in the notion that music is movement, viewing sound not as a secondary response to motion, but as an extension of the body's expressive intent \cite{bang_suspended_2023}.

Closer to the relational framing we develop here, \citet{preisler_when_2026} sonify the interpersonal movement dynamics between two dancers from wearable IMUs, mapping similarity and shared energy to evolving sonic profiles---a dyadic predecessor to the group-scale relational sonification we explore. Closest to our work, \citet{grebel_designing_2026} present C-Battle, a smartphone-based gesture recognition system that sonifies a Hip-Hop movement vocabulary (\textit{Wave}, \textit{Rolls}, \textit{Kickball Change}, \textit{Tutting}) by mapping joint angles to filter parameters and gesture classifications to pre-composed Ableton loops. Across three workshops with novice, amateur, and one expert participant, they found that dancers experienced sonification either as validation of correctly executed movement or as a medium for open-ended exploration, and that system constraints became creative resources. Encypher and C-Battle differ in sensing modality, in how music is produced (generation rather than sonification of pre-composed loops), and in the social framing of the activity (cypher rather than battle). We unpack the sensing comparison in Section~\ref{sec:motion_sensing} and the broader contribution-level differences in the Discussion.

\subsection{Neural Motion-to-Music Generation}
Recent developments in neural audio synthesis have shifted the focus from deterministic mapping to generative systems. Early work by \citet{aggarwal_dance2music_2021} inverted the conventional paradigm---where motion is typically a response to audio---and demonstrated the viability of using deep neural networks to synthesize music directly from dance. Researchers have described generative AI models that support dance-to-music interaction as ``motion-to-music models.''

Recent interactive performances have explored how navigating the latent spaces of generative models can facilitate expressive co-creation. \citet{nabi_embodied_2024} use the RAVE (Real-time Audio Variational Embedding) architecture, leveraging its capacity for low-latency timbre transfer \cite{caillon_rave_2021}. By mapping inertial measurement unit data directly to RAVE's latent space, this approach enables dancers and musicians to co-create in a shared sonic environment. Building on this, \citet{meyer_interactive_2025} replace wearable sensors with computer vision, training lightweight autoencoders to map video input directly to sound in real time. While these systems excel at reactive sound production, they often lack the structural complexity found in traditional musical compositions.

To address musical structure, diffusion-based methods have introduced tailored architectures for aligning motion and audio spaces, achieving higher musical quality and beat alignment scores than previous models \cite{you_momu-diffusion_2024, zhang_dance2music-diffusion_2024}. \citet{you_momu-diffusion_2024} developed a bidirectional contrastive rhythmic VAE (BiCoR-VAE) combined with a diffusion transformer to learn rhythmic structure from vision-based motion input, while \citet{wang_discrete_2025} applied a discrete diffusion model with contrastive learning on the AIST++ dataset to generate dance from music, thus extending the motion--music learning loop in the opposite direction \cite{li_ai_2021}. These systems, however, operate offline: the iterative nature of diffusion struggles to meet the low-latency requirements of live performance.

\textit{Con Moto} \cite{chen_con_2026} addresses this latency--coherence trade-off by leveraging real-time MIDI-based transformer models to translate camera motion data into movement parameters that steer musical output at inference time. Extending beyond individual embodied music-making, it also explores the social and interpersonal dynamics between two dancers within generative loops. By offering configurable levels of agency---from fine-grained control over individual musical voices to abstract mappings that position the system as an autonomous creative partner---\textit{Con Moto} demonstrates that the aesthetic friction between human intent and AI agency can itself become a generative resource. While quantitative work has shown that dyadic motion coordination is reliably measurable and perceptible to observers \cite{velletaz_automatic_2026}, past systems that center the dyadic or individual performer--system relationship leave open the question of how such frameworks might support social dynamics among a group of dancers.

\section{Methods}

Building on this motivation, this work asks: How can generative music support social interactions in dance? To investigate this, we draw on the social interaction dimensions outlined above---social presence, group cohesion, style, quality, facilitation, invitation, and encouragement---to guide the design of shared, embodied experiences through generative music. Different motion-to-music systems afford different forms of participation. We evaluate how various mappings, or interaction styles, facilitate, invite, and encourage social interaction, and assess these interactions in terms of social presence, group cohesion, and quality. By situating participants in embodied, collective experiences, the work explores how AI can mediate not only music generation, but also human-to-human connection.

Previous motion-to-music systems have largely centered on solo performance and fixed stylistic domains. We extend these paradigms toward multi-person interaction, where mappings designed for individual control may no longer be sufficient. New forms of mapping---such as those based on proximity, relative movement, or group configuration---offer opportunities to capture the social dynamics between participants rather than their isolated gestures.

\subsection{First-Person Approach}

Our research follows a first-person co-creation approach with dancers, drawing on Tove Bang's thesis \textit{Designing in conversation with dance and movement practice using first-person methods} \cite{bang_designing_2025}, Stephen Brade's work in human-AI collaborative improvisation \cite{brade_agents_2026}, and the co-design methodology for embodied music steering presented in \textit{Con Moto}.

Bang's thesis explores how to design embodied digital technologies centered on the idea that music is movement. Her autoethnographic methods showcase soma design, movement-based design, and co-design with slowness, humility, and attention to subtle movement qualities, revealing how interaction design can grow out of the lived experience of dancers. Brade defines a framework for interacting with novel AI music generation models through iterative design and personalization. Together, Bang's and Brade's practices suggest that designing interactive systems for dance benefits from centering dancers as co-designers, attending to embodied, tacit knowledge, and exploring how different mapping strategies for music interact with bodily experience.

While the design process is rooted in the first author's own movement practice, we recognize that a single somatic perspective is insufficient for broad system development. Each iteration therefore engages a diverse group of dancers in co-creative sessions to refine motion-to-sound mappings through shared embodied experimentation. Iterative prototyping sessions invite dancers to move with and against the system to understand how the mappings shape their sense of flow, agency, and musicality. Feedback from these sessions guides both technical adjustments and conceptual refinements to the interaction design. This cyclical process grounds system development in lived, bodily experience, ensuring that each layer of technological design remains connected to the dancer's expressive intent and perception.

\subsection{Design Space}

Encypher builds on live music models \cite{team_live_2025}: codec language models in the lineage of AudioLM and MusicLM, in which a Transformer autoregressively predicts discrete audio-codec tokens conditioned on text or audio style prompts. This is a different branch from the symbolic-token transformers---Music Transformer \cite{huang_music_2018} and the Anticipatory Music Transformer \cite{thickstun_anticipatory_2023}---that \textit{Con Moto} steered, and from the offline diffusion models discussed above. That shift trades the note-level control of symbolic models for the timbral richness of studio-quality audio, a trade-off we return to in the Discussion. We care about diversity in two senses: variety in musical output, and the ability to adapt to different performers, cultural perspectives, and group sizes. To reason about how different motion-to-music systems afford different forms of participation, we organize them into four primary modes---structural, reactive, semantic, and textural---each reflecting a distinct pattern of coupling between movement and sound.

\begin{description}
    \item[Structural systems] Mappings that evolve over time, where accumulated input generates layered textures or longer-term structure. For example, Dance2Music-Diffusion \cite{zhang_dance2music-diffusion_2024} and MoMu-Diffusion \cite{you_momu-diffusion_2024} extend music generation using latent diffusion and generate outputs from short audio clips of fixed duration. Similarly, Be the Beat \cite{chang_be_2025} requires a 5-second input to suggest a song.

    \item[Reactive systems] Immediate mappings between movement and sound that allow live improvisation and rhythmic synchronization. Reactive systems create a sense of causal coupling between movement and sound, where dancers expect each gesture to produce an immediate sonic effect. This temporal precision is essential for maintaining rhythmic entrainment and embodied flow, as shown by \citet{nabi_embodied_2024}'s exploration of RAVE latent space through motion. In the domain of live musical improvisation, JAM\_BOT \cite{blanchard_jam_bot_2025} is a real-time symbolic music generation system that enables human-AI jamming in a performance setting.

    \item[Semantic systems] AI models that classify gestures or bodily states to select or generate distinct musical outputs. Be the Beat \cite{chang_be_2025} is a semantic system that captures the genre, energy, and tempo of a dance as prompts to suggest songs from Spotify. Other semantic approaches use language or symbolic prompts to steer music generation \cite{wu_midi-llm_2025, team_live_2025}.

    \item[Textural systems] Mappings that focus on the continuous surface of sound and motion, capturing subtle variations in velocity, pressure, or orientation to shape sonic texture \cite{caillon_rave_2021}. Models like RAVE excel in textural mapping because they allow tone transfer and nuanced sonic response to microgestures, creating a rich moment-to-moment mapping between movement and sound. However, these systems do not model long-term structure, so they are most suited to exploring immediate interactions rather than persistent musical form.
\end{description}

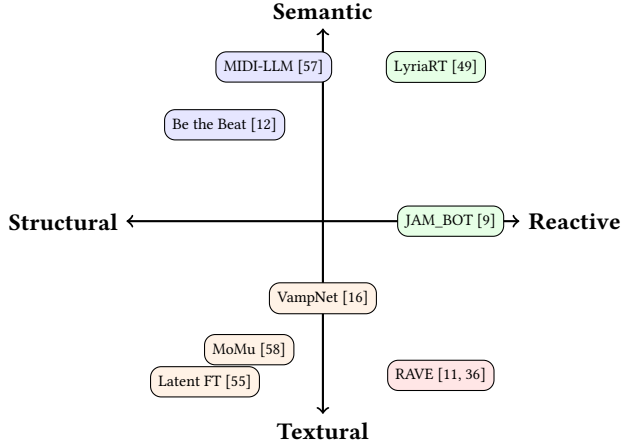
\begin{figure}[htbp]
\centering
\begin{tikzpicture}[xscale=0.65, yscale=0.85]

    \draw[<->, thick] (-4,0) -- (4,0) node[right] {\textbf{Reactive}};
    \node[left] at (-4,0) {\textbf{Structural}};

    \draw[<->, thick] (0,-3) -- (0,3) node[above] {\textbf{Semantic}};
    \node[below] at (0,-3) {\textbf{Textural}};

    \node[draw, fill=blue!10, rounded corners, font=\scriptsize] (BTB) at (-2, 1.5) {Be the Beat \cite{chang_be_2025}};
    \node[draw, fill=blue!10, rounded corners, font=\scriptsize] (LLM) at (-1, 2.4) {MIDI-LLM \cite{wu_midi-llm_2025}};

    \node[draw, fill=green!10, rounded corners, font=\scriptsize] (JB) at (2.6, 0) {JAM\_BOT \cite{blanchard_jam_bot_2025}};
    \node[draw, fill=green!10, rounded corners, font=\scriptsize] (LYR) at (2.3, 2.4) {LyriaRT \cite{team_live_2025}};

    \node[draw, fill=orange!10, rounded corners, font=\scriptsize] (VN) at (0, -1.2) {VampNet \cite{garcia_vampnet_2023}};
    \node[draw, fill=orange!10, rounded corners, font=\scriptsize] (MM) at (-1.5, -2) {MoMu \cite{you_momu-diffusion_2024}};
    \node[draw, fill=orange!10, rounded corners, font=\scriptsize] (LFT) at (-2.4, -2.5) {Latent FT \cite{wang_latent_2026}};

    \node[draw, fill=red!10, rounded corners, font=\scriptsize] (RV) at (2.4, -2.4) {RAVE \cite{caillon_rave_2021, nabi_embodied_2024}};

\end{tikzpicture}
\caption{Classification of AI music systems by temporal response and mapping depth.}
\Description{A two-axis diagram placing AI music systems along axes of Structural--Reactive (horizontal) and Textural--Semantic (vertical).}
\label{fig:system_grid}
\end{figure}

\subsection{Motion Sensing}
\label{sec:motion_sensing}

A central design choice in Encypher is the use of a wide-angle stereo depth camera (the ZED~2i) for multi-person tracking, rather than body-worn sensors such as IMUs or smartphones. This decision is informed both by \textit{Con Moto}, which used the same camera platform successfully for dyadic dance improvisation, and by the specific demands of the cypher context.

Depth-stereo sensing offers three properties that align with our research goals. First, it captures spatial relationships among participants---proximity, clustering, and group formation---in addition to individual joint trajectories, enabling the system to reason about social geometry rather than only individual gesture. Second, it removes the need for wearable hardware, preserving the unencumbered movement vocabulary of Hip-Hop and supporting floor-based styles such as breaking that are incompatible with limb-mounted devices. Third, it is robust enough for ecological deployment: prior work demonstrates sub-10\,ms skeleton capture latency \cite{sinclair_dancegraph_2023}, depth error under 20\,mm at distances up to 7\,m \cite{abdelsalam_depth_2024}, and superior limb-tracking performance of the ZED~2i over comparable sensors \cite{aharony_comparative_2023, sosa-leon_evaluating_2022}. The ZED~2i's wider field of view and longer depth range make it particularly well-suited to group-scale capture in varied lighting, and prior dance applications \cite{balafoutis_tradance_2023} further support the feasibility of camera-based sensing for movement analysis.

These properties contrast directly with the design choices of C-Battle \cite{grebel_designing_2026}, which straps iPhones to dancers' wrists and shins. While this enables fine-grained joint-angle capture, the form factor rules out cypher contexts and excludes large parts of the Hip-Hop and street dance vocabulary. Floor-based styles such as breaking---windmills, freezes, headspins---cannot be performed with phones strapped to limbs. Arm-heavy styles such as locking and waacking are similarly disadvantaged: the added weight on the wrists affects movement quality, and the device itself constrains rotational range at the joint. Our initial co-design sessions with breakers confirmed that a hands-free, floor-inclusive sensing surface was a prerequisite for the system to be welcome in their practice. Camera-based collective sensing accepts a different set of trade-offs (multi-angle setup, lighting sensitivity, calibration) in exchange for a sensing surface that does not encode a theoretical judgment about which dance styles belong in the cypher.

\paragraph{Deployment Configuration and Observed Limits}

In our installations, the ZED~2i is mounted at 2.2\,m, angled 30$^\circ$ downward, running the ZED SDK's \textsc{Body\_34} skeleton tracking on HD720 input with neural depth at 30\,fps on a laptop with an RTX 4090 GPU, with all joints in world-frame metres so that group-scale metrics such as inter-participant distance and circularity need no unit conversion. The system reliably detected up to 15 people, but joint tracking degraded when participants stood behind one another, confirming \citet{grebel_designing_2026}'s concern about occlusion in cypher formations. The SDK's multi-camera Fusion module improved occlusion robustness but halved throughput on our hardware. The single-camera setup was sufficient for groups of up to six---the size used in our user study---and we treat scaling to larger groups as a deployment, rather than design, problem for future work.

\section{System Design}

\subsection{System Overview}

The design of the Encypher system investigates how a live music model can serve as a creative, collaborative dance-music partner. While an earlier MIDI-based prototype allowed for fine-grained control of music generation, it was limited in the complexity of musical textures that could be generated in real time. The system design, built on Lyria RealTime (LyriaRT), was organized around two key considerations: musical richness sufficient to invite dancing, and a control surface coarse enough that a group of dancers could shape the output collectively without any one person needing precise individual control.

The system is organized as a real-time control loop that couples embodied movement features and music generation in the audio domain. Encypher consists of three primary components: (1) a web interface that handles LyriaRT API calls, music generation, and visualization; (2) a depth camera and feature-extraction process that estimates each dancer's position and activity level from joint trajectories; and (3) a Max/MSP patch that serves as a central hub for communication between the Windows laptop, the web interface, and the automated lighting controls.

The generative model uses LyriaRT, ``a live music model that generates an uninterrupted stream of music and responds continuously to user input'' \cite{team_live_2025}. MagentaRT and LyriaRT are both developed by the Lyria Team at Google DeepMind. Compared to MagentaRT, which is open-weights and trained primarily on Western instrumental music, LyriaRT is trained on a significantly larger and more diverse dataset including high-fidelity studio recordings. These models continuously generate audio in two-second chunks with 10 seconds of past audio context.

Motion data is streamed from the camera analysis process to Max/MSP via OSC, following the earlier prototype. The analysis process extracts a compact set of features---spatial location on stage, activity level, and zone occupancy---which are transmitted at interactive rates. Within Max/MSP, these features are filtered, smoothed, and quantized to reduce jitter and to convert continuous movement into musically meaningful control signals.

\subsection{Embodied Collaborative Steering}

Rather than exploring multiple configurable levels of agency as in \textit{Con Moto}, Encypher extends the autonomous partnership model by mapping collective movement qualities directly to text prompt weights in LyriaRT. We co-designed the following movement-to-music mappings with breakers in the local community.

\paragraph{Energy-Driven Prompt Steering}

LyriaRT's architecture supports weighted prompts, allowing continuous control over the generation process. Building on \textit{Con Moto}'s mapping of movement velocity to MIDI velocity, Upper Body Center of Mass Amplitude (COMAmp) serves as our primary movement feature---reflecting how much participants are ``grooving,'' a low-energy rocking or swaying motion that does not require stepping or arm gestures. Focusing on groove enables broader participation across different Hip-Hop styles and skill levels. Concretely, COMAmp measures how much a participant's upper-body center of mass oscillates over a rolling two-second window, computed as the L2 norm of the standard deviations of the torso's X and Y world positions---combining horizontal bounce and lateral rock into a single 0--5 scalar value, smoothed via exponential moving average ($\alpha$ = 0.3) to reduce jitter. Each dancer's COMAmp is then mapped to the weight of a prompt representing their assigned musical element, and these weights sum across all participants, creating a collective steering signal where the aggregate energy of the group shapes the generation in real time.

\paragraph{Temperature and Top-K Control}

To support group dynamics, the system couples generation diversity with group formation. LyriaRT's temperature and top-K parameters control randomness and diversity in music generation; experiments revealed that high values produce chaotic, incoherent results that dancers find difficult to move with. Rather than requiring manual parameter tuning, we designed the system to measure group circularity---how many of the six stationary zones are occupied---and map this directly to these parameters. When dancers form a tight cypher (high zone occupancy), the music becomes more coherent and predictable; when scattered (low occupancy), the system embraces randomness. This coupling creates a feedback loop: better musical coherence encourages dancers to move closer together, reinforcing the social formation itself. The circularity score is smoothed via exponential moving average to reduce jitter and create stable musical transitions.

\subsection{Co-Design Process}

The development of Encypher followed an iterative co-design methodology driven by weekly sessions with a local breaking crew, Imobilare,
during their Monday open training sessions. Rather than isolating the system in a laboratory, the first author integrated the system into the dancers' existing practice, setting up in the corner and inviting engagement as part of their regular freestyle session. The co-design process prioritized the dancers' embodied experience and social context over controlled variables, treating the cypher's low-stakes, vulnerable environment---a space for learning, play, and technique improvement---as essential to understanding how the system could support authentic social interaction.

\paragraph{Calibrating Movement Vocabulary}

Early prototypes assigned each dancer a discrete musical element, creating confusion about role assignment and entry points. To resolve this, we co-designed six stationary zones arranged in a circle, each pre-labeled with a musical element. Defined zones reduced cognitive overhead and naturally supported a cypher formation. However, energy control proved critical: when we adapted the limb-velocity mapping from \textit{Con Moto}, dancers found it unnatural. Breakers groove with body sway and rocking motion rather than sharp limb movements. We shifted to COMAmp, capturing movement that does not require stepping or arm gestures. The insight---that interaction models must be calibrated to specific movement vocabularies---directly shaped the embodied collaborative steering design.

\paragraph{Addressing Latency Through Co-Designed Affordances}

LyriaRT's two-second generation latency fundamentally challenged the sense of causal coupling dancers expected. Rather than treating this as a technical failure, we co-designed a ``charging'' mechanic: dancers build energy on their zone, and once it reaches a threshold, the music responds. This transformed a constraint into an affordance---the delay became intentional, teaching users that collective presence precedes musical response.

\paragraph{Genre Coherence and Preset Configurations}

Experiments with arbitrary musical prompts (e.g., six very different genres) produced outputs that failed to represent the prompts faithfully. Dancers observed that coherence increased when the system committed to a single genre and stacked compatible instruments (e.g., Breaking + Snares + Keys + Bass Guitar + Guitar + Drums). We implemented a preset system to save pleasing combinations as JSON configurations, shifting the design from granular parameter tuning to curated, genre-aware generation.

\paragraph{From Individual to Collective Control}
Initial designs explored temperature and top-K parameters as fine-grained control mechanisms, with arm velocity mapped to top-K and leg velocity to temperature---separate from COMAmp and intended to give dancers more individual control over the music. In practice, however, neither parameter alone produced noticeable musical change, and combining both introduced unpredictable variation that dancers found disruptive rather than expressive, making the music feel chaotic and less suited to dancing.
Iterating on these failed mappings led to the idea of circularity: rather than giving individuals more control, we shifted focus toward bringing people together spatially. This design shift also surfaced a methodological insight---testing with dancers who already knew each other obscured whether the system could genuinely facilitate social connection, since existing relationships masked the effect.
Dancers also proposed tempo control, a recurring design desire across sessions. However, BPM steering proved unreliable with text-conditioned generation: certain genre and instrument combinations resisted tempo constraints, producing musically inconsistent results. We treated this as an engineering limitation outside the scope of the current work and prioritized other interaction design questions.

\paragraph{Community Constraints and Research Pivot}

Over five weeks, breaker dancers provided extensive technical feedback, surfacing bugs and suggesting interaction improvements. Yet this feedback could not answer the core research question: \textit{how can generative music support social interaction among people who do not know each other?} Established communities already possess strong social dynamics; the system became a novelty add-on rather than a social bridge. This realization changed the study design: designing \textit{with} an established community and designing \textit{for} social connection among strangers are fundamentally different challenges. The decision to shift focus to novices and unacquainted participants---informed by this co-design experience---became essential to addressing the research question.

\section{Movement Class User Study}

An opportunity arose to conduct a formal user study with participants from a non-expert population. An introductory movement class at MIT
provided access to students with minimal dance experience and no pre-existing social ties---ideal conditions for testing whether Encypher could facilitate connection among strangers.

\subsection{Participants}

We recruited 11 adult participants from the course. %
Participation was voluntary, with no impact on course standing or grade, and informed consent was obtained from each participant under an IRB-approved protocol.

\subsection{Setup}

The study was conducted in a professional dance studio with black marley floors and full-length mirrors, where the class usually met. The space was outfitted with the Encypher installation: a depth camera mounted to track participants in the shared space, a laptop running the web visualizer and LyriaRT API-calling pipeline, a large-screen projection of the six musical zones arranged in a circle, and stereo audio playback. From the second group onward, each zone was also marked physically on the floor with tape (an ``X'') and pre-labeled with a musical element (e.g., drums, bass, keys). We recorded video of the sessions for later qualitative analysis of social cues such as eye contact, approach/retreat behavior, and group formation.

\subsection{Procedure}

\begin{figure}
    \centering
    \includegraphics[width=\linewidth]{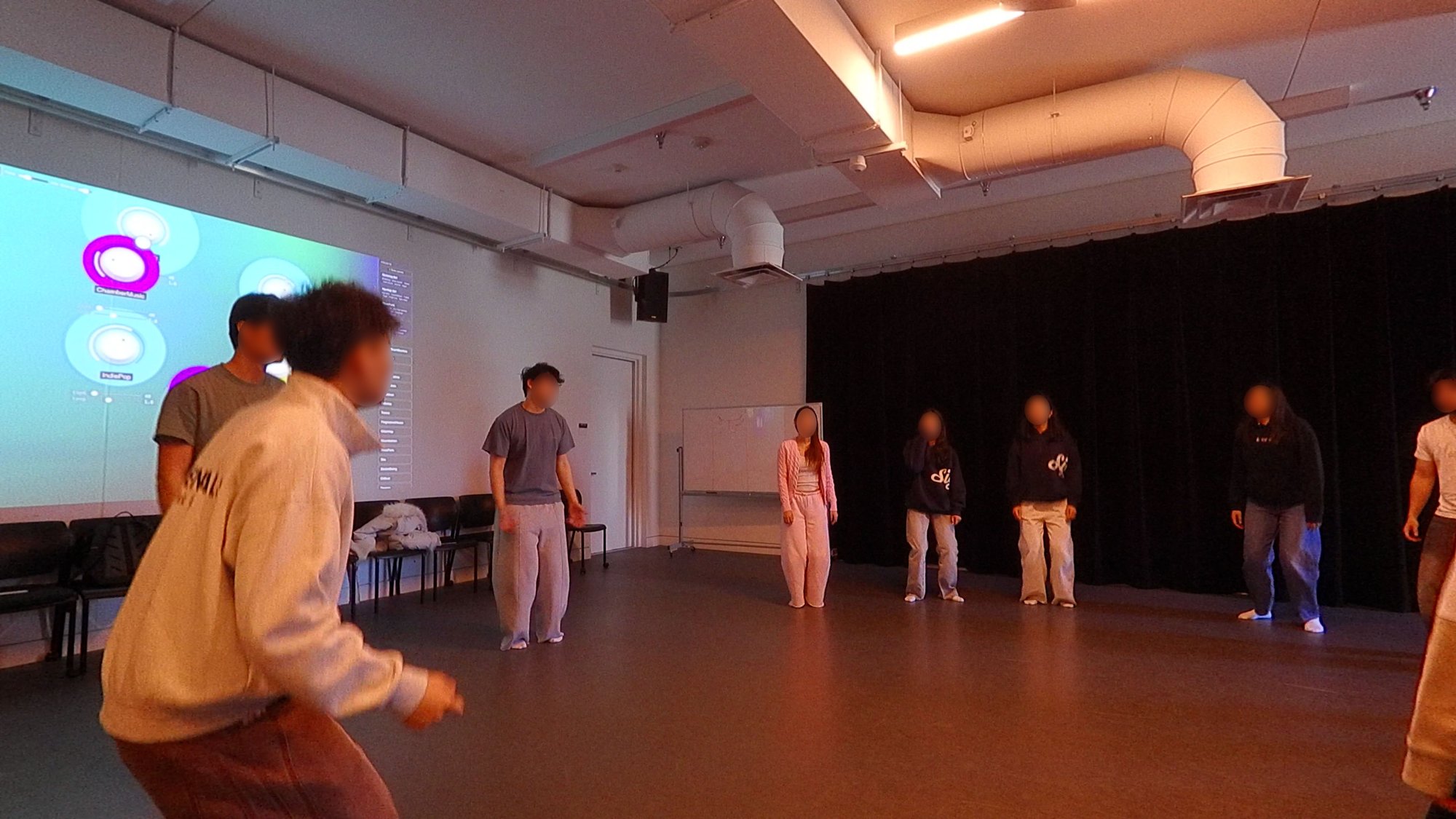}
    \caption{Warmup activity at the start of the user study session, where participants engaged in collective bouncing to acclimate to the movement vocabulary and the shared space.}
    \Description{Participants standing in a loose group in a dance studio, bouncing together during a warmup; faces are blurred.}
    \label{fig:warmup}
\end{figure}

The one-hour session followed a structured agenda. After signing consent forms, the class began with a collective warmup activity (Figure~\ref{fig:warmup}) that was part of the class's normal routine, allowing participants to acclimate to the movement vocabulary of ``grooving'' and to one another. Participants were then divided into two groups (A: n=6, B: n=5), with each group taking turns dancing with the Encypher system while the others observed from outside the cypher.

Each group danced for 5--10 minutes, after which the group that had just danced completed a paper questionnaire while the next group entered the cypher. Group A danced an additional round at the end. The questionnaire (Appendix~\ref{app:questionnaire}) used a 7-point Likert scale across nine dimensions of social and personal experience (engagement, control, enjoyment, replay, social presence, collaboration, invitation, encouragement, spatial awareness), followed by open-ended reflection prompts about connection, agency, hesitation, and the role of the marked spots.

\begin{figure}
    \centering
    \includegraphics[width=\linewidth]{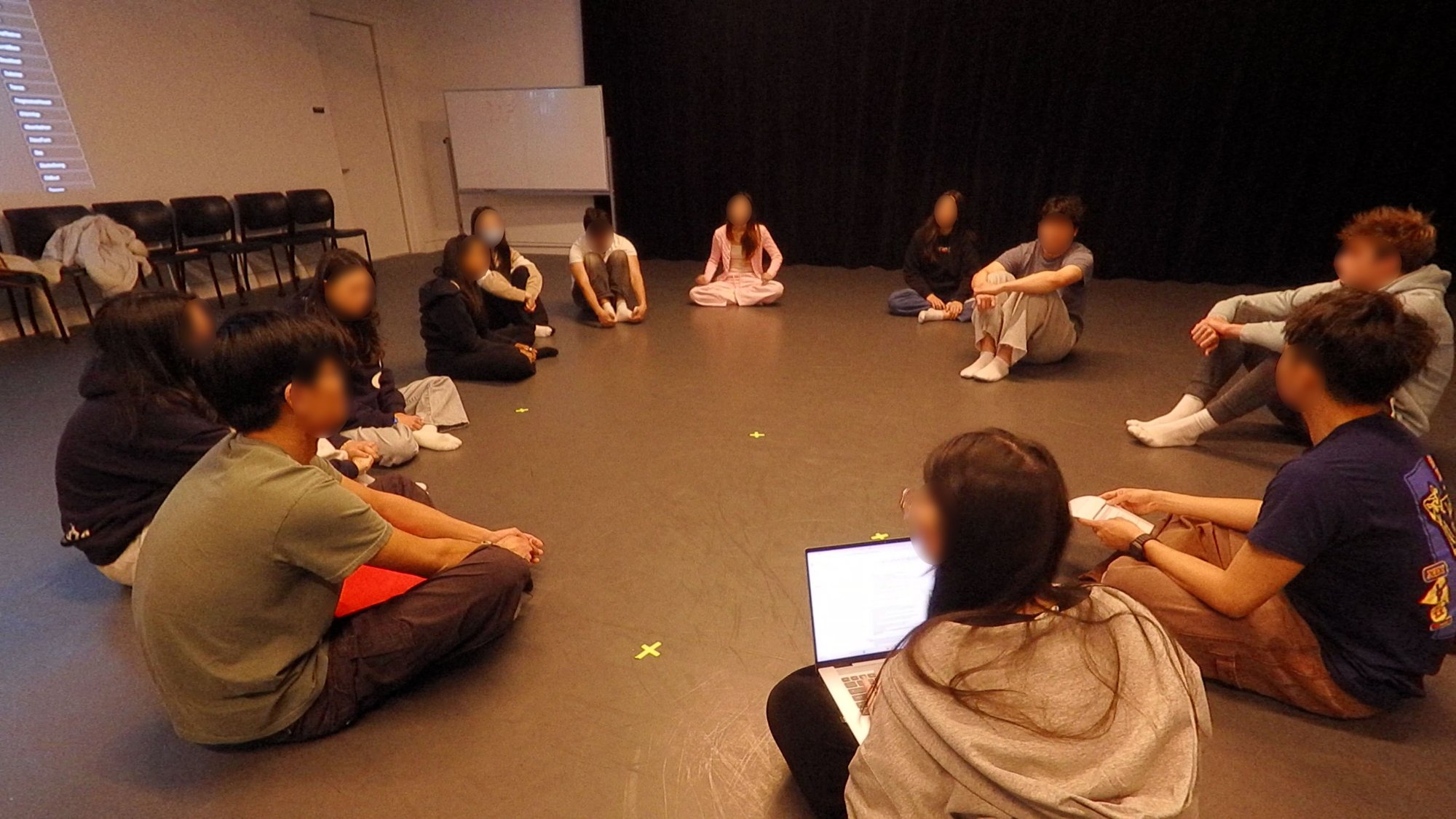}
    \caption{Post-session discussion circle, where participants reflected on their experiences with the Encypher system and shared their perceptions of agency, social presence, and group dynamics.}
    \Description{Participants seated in a circle on a dance studio floor during a group discussion; faces are blurred.}
    \label{fig:discussion_circle}
\end{figure}

The session concluded with a 20-minute facilitated group discussion (Figure~\ref{fig:discussion_circle}). Discussion prompts focused on collective experience rather than individual responses, encouraging participants to build on one another's accounts. Topics included moments of felt connection, perceptions of agency over the music, hesitations about joining, and the influence of spatial markers on movement and attention.

\begin{figure*}[htbp]
    \centering
    \begin{tikzpicture}
        \begin{axis}[
            xbar stacked,
            width=0.85\textwidth,
            height=10cm, %
            inner sep=0pt,
            xmin=-100, xmax=100,
            xtick={-100,-50,0,50,100},
            xticklabels={100\%, 50\%, 0, 50\%, 100\%},
            xlabel={Percentage of Responses (Disagree $\leftarrow \rightarrow$ Agree)},
            ytick=data,
            symbolic y coords={
                {Spatial Awareness},
                {Encouragement},
                {Invitation},
                {Collaboration},
                {Social Presence},
                {Replay},
                {Enjoyment},
                {Control},
                {Engagement}
            },
            axis x line=bottom,
            axis y line=left,
            enlarge y limits=0.1,
            legend entries={Strongly Disagree (1-2), Disagree (3), Neutral (4), Agree (5-6), Strongly Agree (7)},
            legend style={at={(0.5,-0.15)}, anchor=north, legend columns=3, font=\small, draw=none},
            area legend
        ]

        \addlegendimage{fill=red!70!black}
        \addlegendimage{fill=red!40}
        \addlegendimage{fill=gray!30}
        \addlegendimage{fill=blue!40}
        \addlegendimage{fill=blue!70!black}

        \addplot[fill=gray!30, forget plot] coordinates {
            (-22.8,{Engagement}) (-4.5,{Control}) (-13.7,{Enjoyment}) (-13.7,{Replay})
            (-13.7,{Social Presence}) (-4.5,{Collaboration}) (-9.1,{Invitation})
            (-9.1,{Encouragement}) (-0.0,{Spatial Awareness})
        };

        \addplot[fill=red!40, forget plot] coordinates {
            (-0.0,{Engagement}) (-27.3,{Control}) (-0.0,{Enjoyment}) (-0.0,{Replay})
            (-9.1,{Social Presence}) (-0.0,{Collaboration}) (-18.2,{Invitation})
            (-9.1,{Encouragement}) (-9.1,{Spatial Awareness})
        };

        \addplot[fill=red!70!black, forget plot] coordinates {
            (-0.0,{Engagement}) (-0.0,{Control}) (-0.0,{Enjoyment}) (-9.1,{Replay})
            (-9.1,{Social Presence}) (-9.1,{Collaboration}) (-0.0,{Invitation})
            (-0.0,{Encouragement}) (-0.0,{Spatial Awareness})
        };

        \addplot[fill=gray!30, forget plot] coordinates {
            (22.8,{Engagement}) (4.5,{Control}) (13.7,{Enjoyment}) (13.7,{Replay})
            (13.7,{Social Presence}) (4.5,{Collaboration}) (9.1,{Invitation})
            (9.1,{Encouragement}) (0.0,{Spatial Awareness})
        };

        \addplot[fill=blue!40, forget plot] coordinates {
            (45.5,{Engagement}) (63.6,{Control}) (63.6,{Enjoyment}) (54.5,{Replay})
            (36.4,{Social Presence}) (63.6,{Collaboration}) (54.5,{Invitation})
            (63.6,{Encouragement}) (54.5,{Spatial Awareness})
        };

        \addplot[fill=blue!70!black, forget plot] coordinates {
            (9.1,{Engagement}) (0.0,{Control}) (9.1,{Enjoyment}) (9.1,{Replay})
            (18.2,{Social Presence}) (18.2,{Collaboration}) (9.1,{Invitation})
            (9.1,{Encouragement}) (36.4,{Spatial Awareness})
        };

        \end{axis}
    \end{tikzpicture}
    \caption{Diverging stacked bar chart showing participants' (n=11) reported experience of the Encypher system across dimensions of social interaction and personal experience.}
    \Description{Diverging stacked horizontal bar chart. Nine rows, from Engagement at the top to Spatial Awareness at the bottom, each showing the percentage of participants who strongly disagreed, disagreed, were neutral, agreed, or strongly agreed. Agreement dominates every row; Control and Invitation have the largest disagree portions.}
    \label{fig:likert_diverging}
\end{figure*}
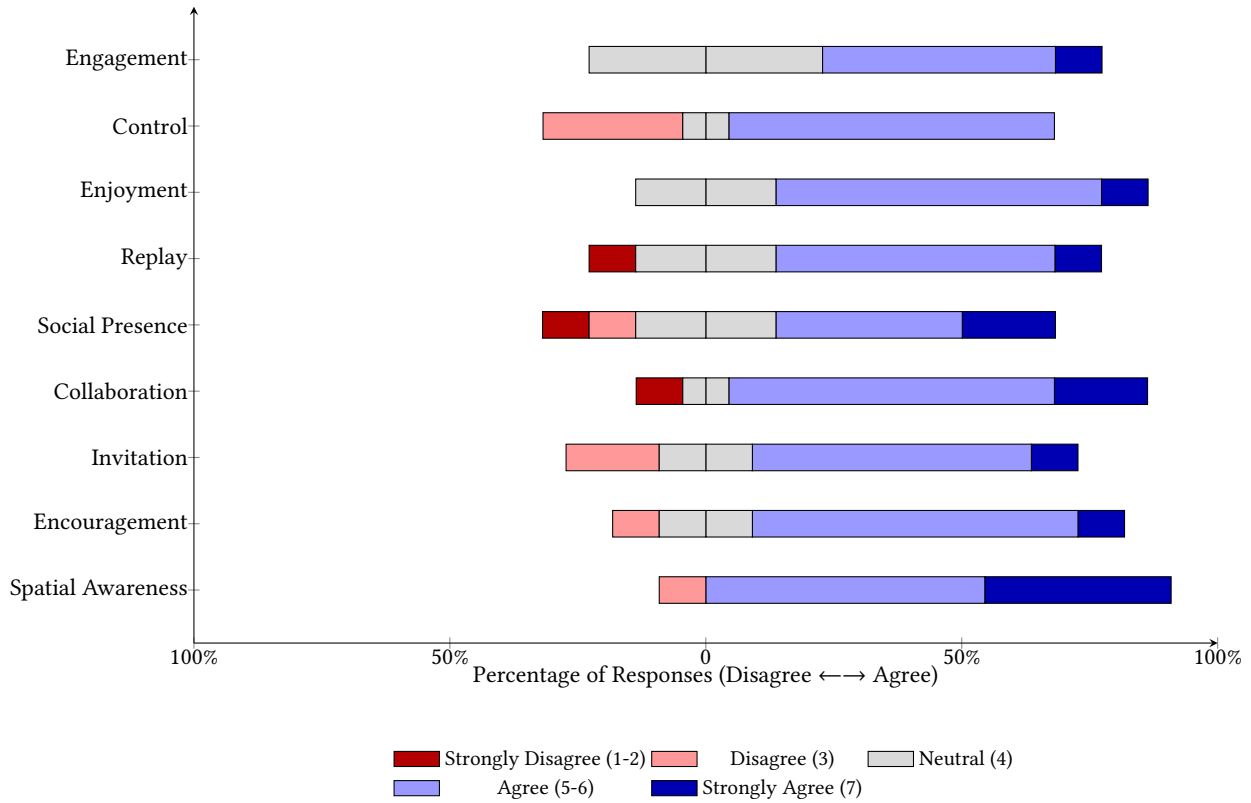

\subsection{Quantitative Analysis}

The quantitative evaluation was based on a 7-point Likert scale (1: Strongly Disagree, 7: Strongly Agree) across nine dimensions of social choreography and interaction quality. The items operationalize the framework in Table~\ref{tab:social_interaction_framework}: social presence maps directly; collaboration to group cohesion; invitation and encouragement to their namesakes; spatial awareness to facilitation; and engagement, enjoyment, and replay to interaction quality. Control is an addition specific to Encypher's design goal of shifting agency from the individual to the group, and interaction style was assessed through video and open-ended responses rather than rated. The full questionnaire is provided in Appendix~\ref{app:questionnaire}.

The quantitative results, visualized in Figure~\ref{fig:likert_diverging}, are positive but moderate. No participant disagreed on \textbf{Engagement} or \textbf{Enjoyment}, but agreement was 54.5\% and 72.7\% respectively, with the remainder neutral; Engagement drew the largest neutral share of any item (45.5\%). \textbf{Spatial Awareness} performed most strongly (90.9\% agreement, 36.4\% strongly agree), the highest ``strongly agree'' rate of any metric, suggesting that the marked zones supported participation. \textbf{Collaboration} reached 81.8\% agreement (18.2\% strongly agree) and \textbf{Encouragement} 72.7\%, consistent with the design hypothesis that generative music can foster shared agency and group cohesion among unacquainted participants.

A more nuanced distribution emerged for \textbf{Social Presence}, where 18.2\% of participants disagreed (9.1\% strongly disagreed) and 27.3\% were neutral or mixed. While the majority (54.6\%) still agreed that the system increased awareness of others, this dimension showed the most variability. We interpret this divergence as evidence of the screen-as-mediator dynamic discussed in our qualitative analysis: the visualization gave participants a shared object of attention that lowered the social barrier, but for some this came at the cost of direct interpersonal awareness. \textbf{Control} similarly showed weaker agreement (63.6\% agreement, 0\% strongly agree, 27.3\% disagree), reflecting the system's intentional design choice to prioritize collective steering over individual agency---a trade-off that successfully shifted the experience toward shared agency, though some participants found the resulting indirectness frustrating. \textbf{Replay} (63.6\% agreement, 9.1\% strongly disagree) and \textbf{Invitation} (63.6\% agreement, 18.2\% disagree) round out the moderate-positive responses, suggesting room for refinement in how the system signals turn-taking and ongoing engagement.

These numbers should be read cautiously. The study was conducted within a credit-bearing classroom environment where the first author was a guest instructor, which may have inclined students toward more positive feedback. The novelty effect of interacting with high-fidelity neural audio for the first time likely also skewed engagement and enjoyment scores upward. To address these limitations, we cross-referenced quantitative responses against open-ended qualitative reflections (Section~\ref{sec:qualitative}), allowing specific technical frustrations and social dynamics to surface alongside the headline numbers.

\begin{figure*}[t]
    \centering
    \begin{subfigure}[t]{0.32\textwidth}
        \centering
        \includegraphics[width=\linewidth]{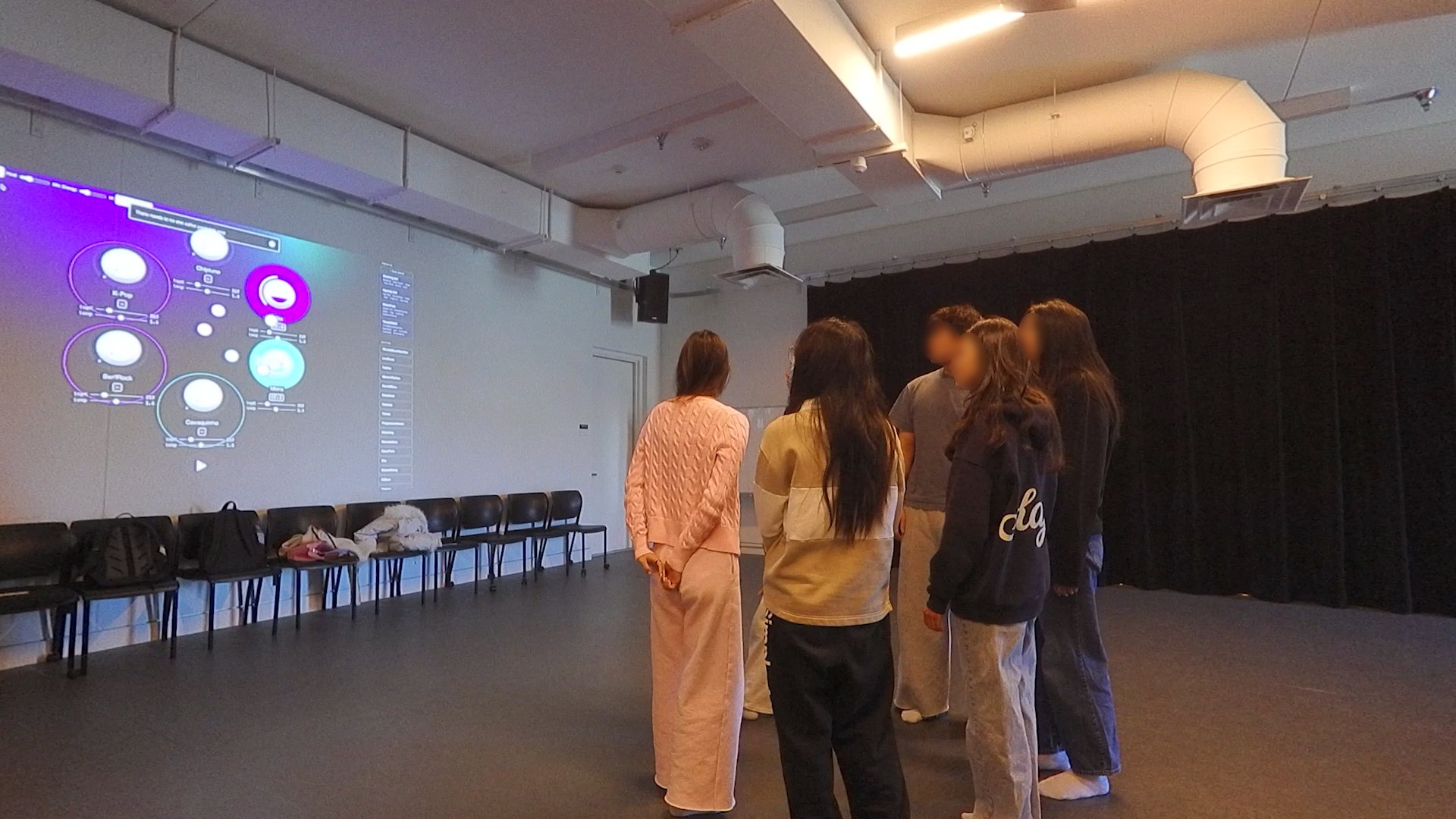}
        \caption{Round 1 (Group A)}
        \label{fig:group1}
    \end{subfigure}
    \hfill
    \begin{subfigure}[t]{0.32\textwidth}
        \centering
        \includegraphics[width=\linewidth]{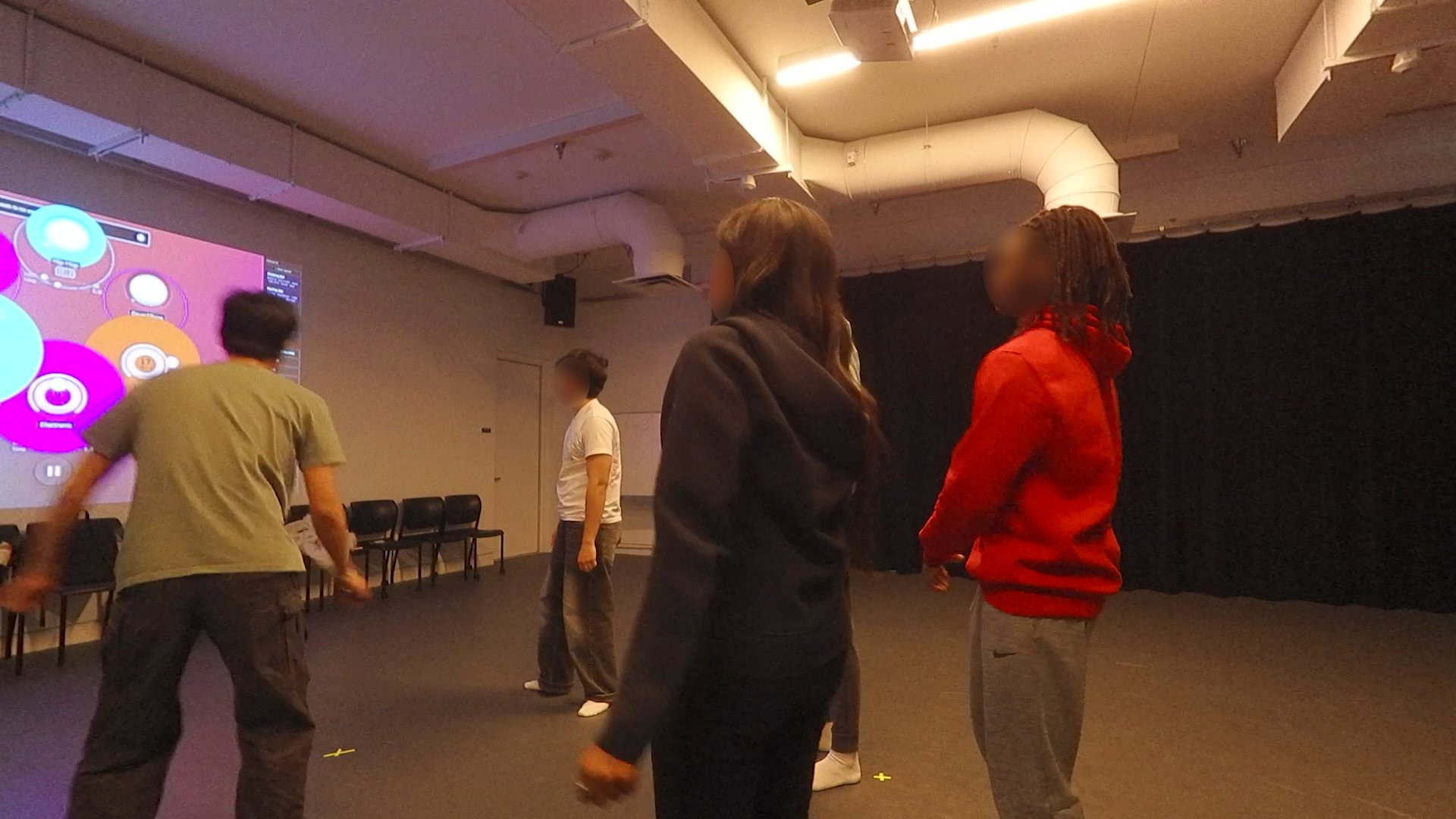}
        \caption{Round 2 (Group B)}
        \label{fig:group2}
    \end{subfigure}
    \hfill
    \begin{subfigure}[t]{0.32\textwidth}
        \centering
        \includegraphics[width=\linewidth]{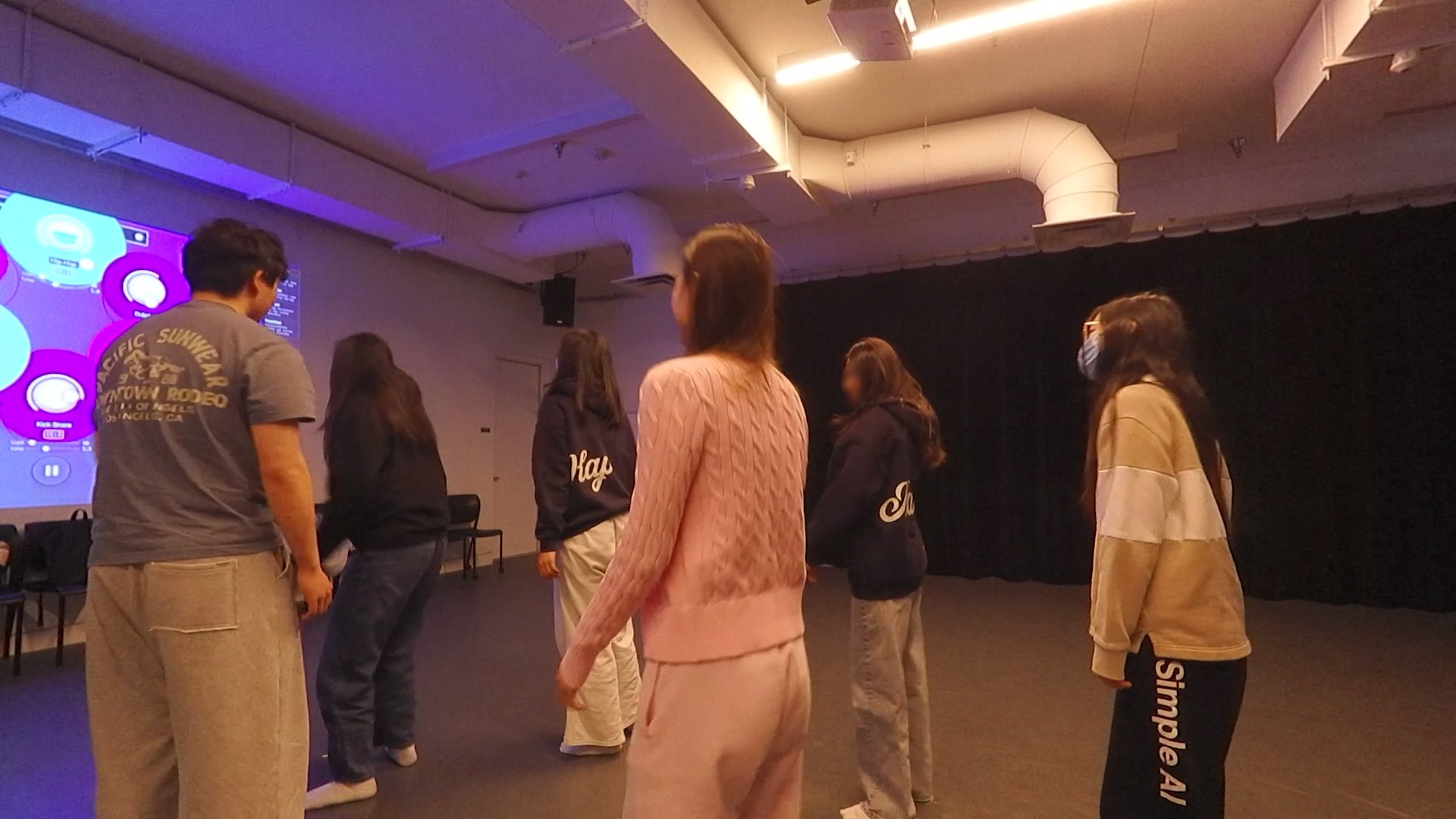}
        \caption{Round 3 (Group A again)}
        \label{fig:group3}
    \end{subfigure}
    \caption{The three rounds of the user study (Group A, Group B, Group A again) dancing with the Encypher system, with the visualization screen visible in the background showing real-time feedback of zone occupancy and energy charging.}
    \Description{Three photographs side by side, one per round, of five or six people dancing in a studio with a projected screen of six colored zones visible behind them; faces are blurred.}
    \label{fig:groups_with_screen}
\end{figure*}

\begin{figure*}[t]
    \centering
    \begin{subfigure}[t]{0.32\textwidth}
        \centering
        \includegraphics[width=\linewidth]{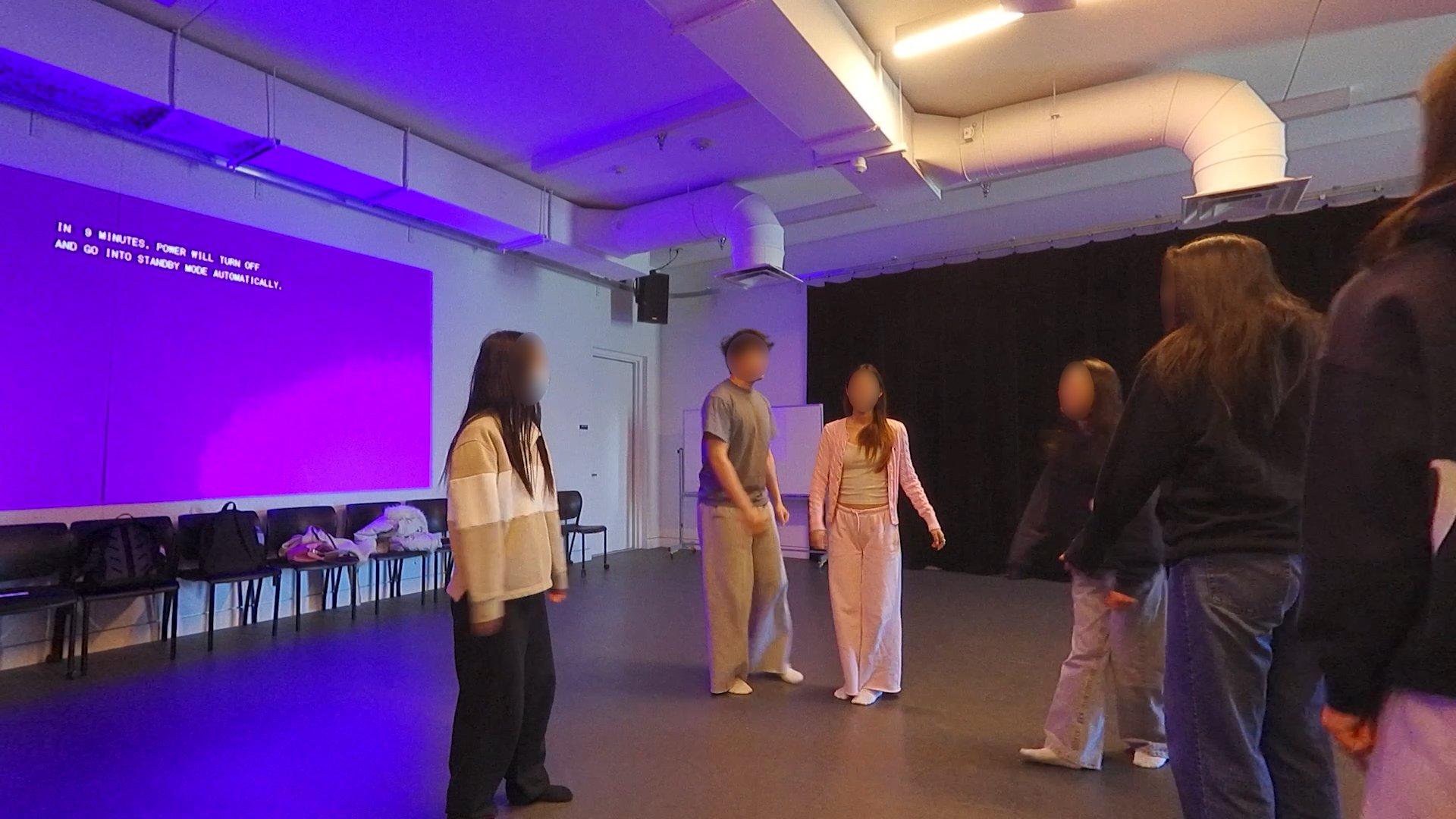}
        \caption{Round 1 (Group A)}
        \label{fig:group1_noscreen}
    \end{subfigure}
    \hfill
    \begin{subfigure}[t]{0.32\textwidth}
        \centering
        \includegraphics[width=\linewidth]{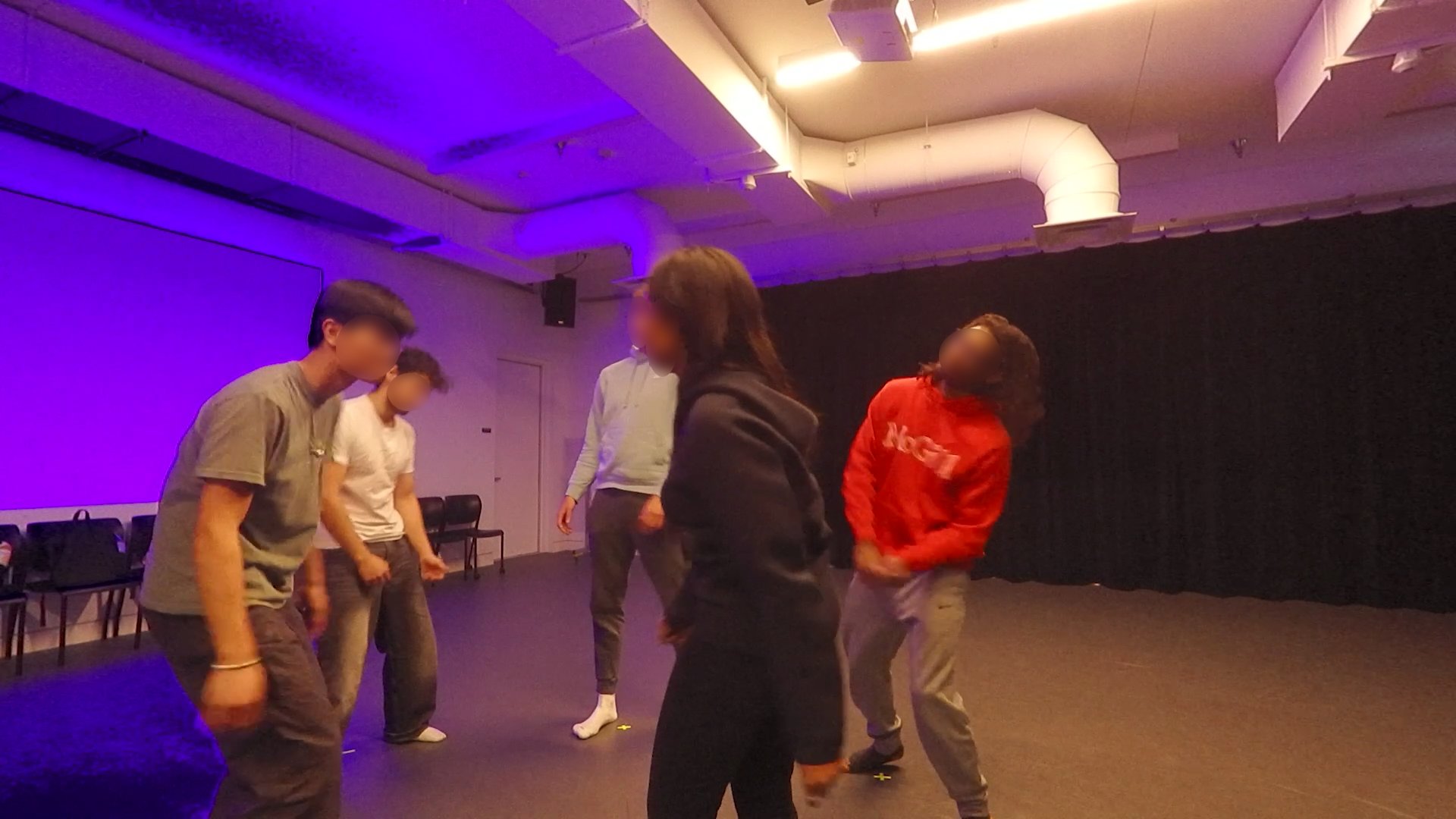}
        \caption{Round 2 (Group B)}
        \label{fig:group2_noscreen}
    \end{subfigure}
    \hfill
    \begin{subfigure}[t]{0.32\textwidth}
        \centering
        \includegraphics[width=\linewidth]{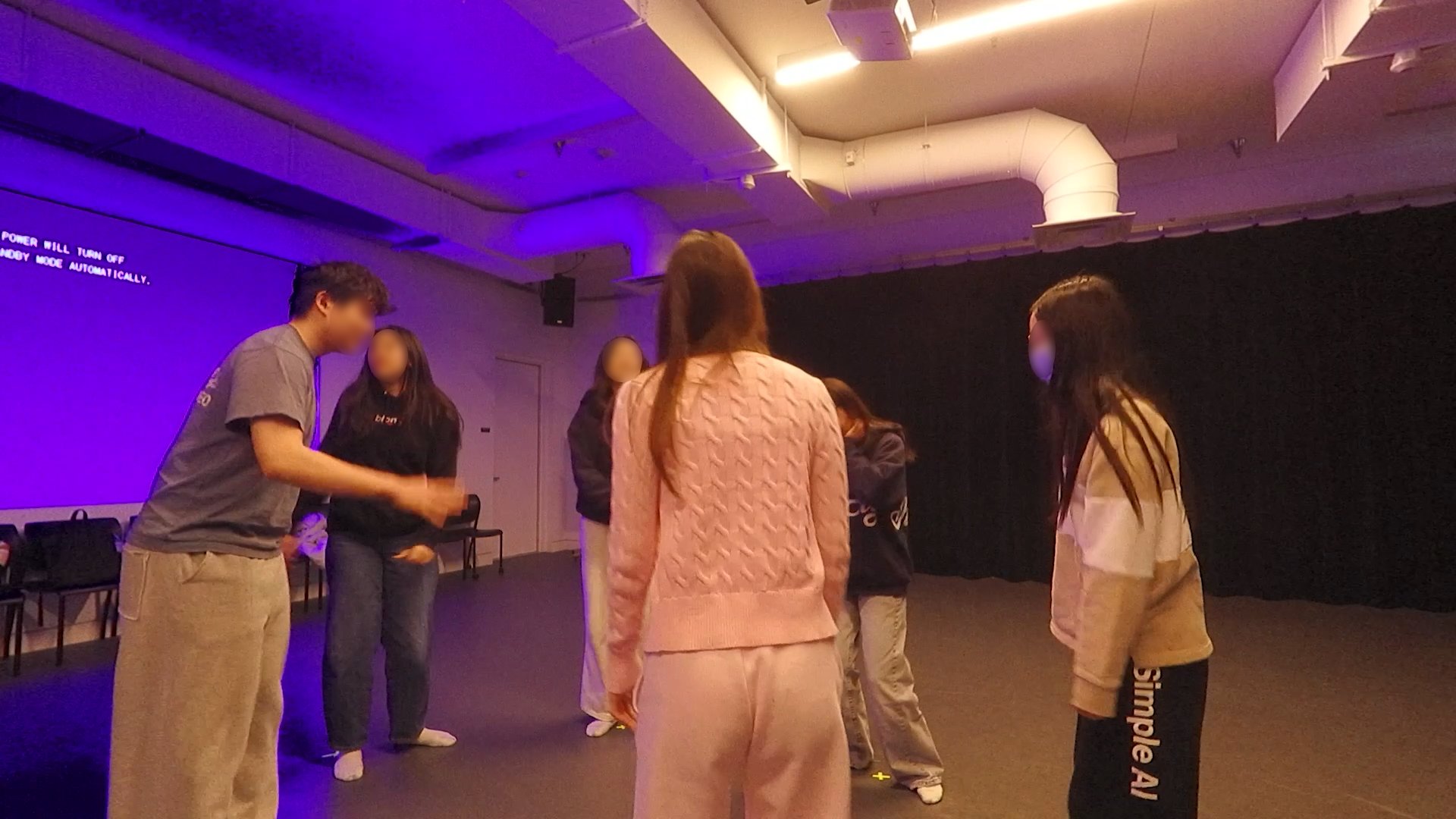}
        \caption{Round 3 (Group A again)}
        \label{fig:group3_noscreen}
    \end{subfigure}
    \caption{The three rounds of the user study dancing without the visualization screen, focusing attention on each other and the embodied experience rather than the interface.}
    \Description{Three photographs side by side, one per round, of people dancing in the same studio with the projection turned off, oriented toward one another; faces are blurred.}
    \label{fig:groups_no_screen}
\end{figure*}

\subsection{Qualitative Analysis: Themes in Collaborative Interaction}
\label{sec:qualitative}

While the quantitative data showed high engagement, the open-ended responses added detail about cypher dynamics, social presence, and the limits of fixed-node interaction. Four primary themes emerged from participants' reflections and our observations, followed by the body-map responses, each illustrating how the system shaped (or failed to shape) shared agency and social presence.

\paragraph{Emergent Non-Verbal Alignment}
Participants described moments of social presence emerging through embodied attunement---noticing one another's movements, mirroring rhythm and energy, and stepping in and out of the cypher in response to visual rather than verbal cues.
\begin{quote}
\textit{``The guy across from me rocked invitingly and I tried to match his movements and energy.''} (P5)
\end{quote}
\begin{quote}
\textit{``When I looked and saw others swinging, I swang [along with them].''} (P8)
\end{quote}
\begin{quote}
\textit{``I noticed the change when stepping out and in of the circle. This made me want to move more.''} (P3)
\end{quote}

\paragraph{Shared Agency as Collective Orchestration}
A key design goal of Encypher was to shift the felt locus of agency from the individual performer to the group. Participants described this shared agency directly, perceiving the music as responding to the room and the collective rather than as a simple one-to-one trigger of their own movement.
\begin{quote}
\textit{``It felt like the music was responding to the group. When the energy was good, it felt like the music was better.''} (P1)
\end{quote}
\begin{quote}
\textit{``I feel like everyone plays a role in creating the music and I really enjoy dancing to the music.''} (P11)
\end{quote}
\begin{quote}
\textit{``Still, when [the sound] changed, I felt like I was stepping in as an instrument.''} (P3)
\end{quote}

\paragraph{The ``X'' as Both Clarity and Constraint}
The marked spots in the space (the ``X'' nodes) created a tension between spatial legibility and the freedom of movement inherent to dance.

\begin{quote}
    As clarity: \textit{``Having marked spots kept people in more structured locations... it gave me clarity on where I could go.''} (P1)
\end{quote}
\begin{quote}
    As constraint: \textit{``I like moving around the space, and the need to stay on the X was quite limiting for me.''} (P9)
\end{quote}
\begin{quote}
    On social territoriality: \textit{``People seemed to own the places they charged up... I felt like I shouldn't move out of the spot where I controlled the sound.''} (P5)
\end{quote}

\begin{figure*}
    \centering
    \includegraphics[width=1\linewidth]{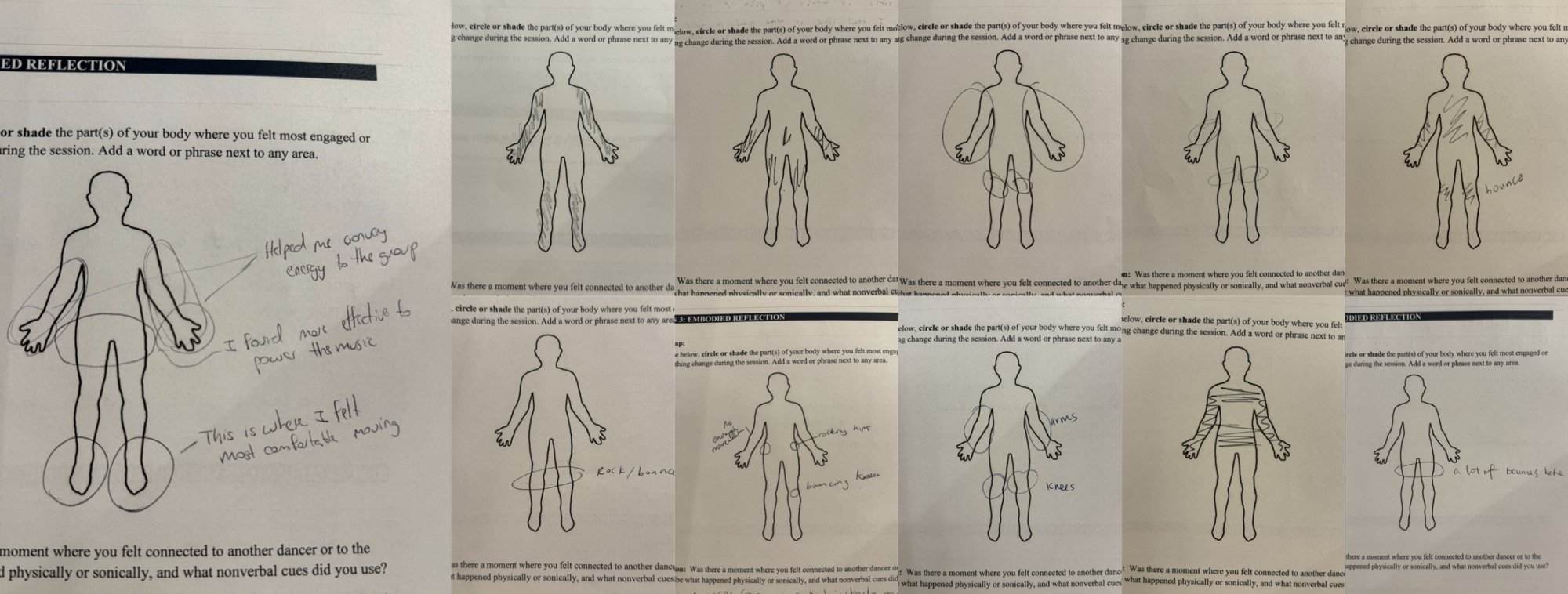}
    \caption{Aggregated body-map responses from participants (n=11), highlighting the body regions where participants reported feeling most engaged or noticing change during the session.}
    \Description{Front and back outlines of a human body with shaded regions showing where participants reported feeling most engaged or noticing change; darker shading marks regions reported by more participants.}
    \label{fig:bodymap}
\end{figure*}
\paragraph{The Screen as Social Mediator}

Comparing sessions with the visualization screen (Figure~\ref{fig:groups_with_screen}) and without it (Figure~\ref{fig:groups_no_screen}) revealed an unexpected role for the visual interface. With the screen, participants frequently looked at the visualization rather than at each other---tracking their avatar's position, watching their charging meter, and checking which zone they occupied. While this might seem counterproductive for a system designed to foster social connection, it actually lowered the social barrier for unacquainted participants. Rather than forcing immediate eye contact or bonding, the screen gave them a shared object of attention---a ``game'' to play together---which assisted their entry into the group dynamic. Figure~\ref{fig:group1} shows how participants gathered close together and oriented toward the screen as they noticed their avatars moving when they started playing.

Without the screen, the social dynamic shifted notably. Initial awkwardness emerged as participants no longer had the visualization as a reference point and had to look at one another directly. Figure~\ref{fig:group1_noscreen} shows this transition: without the visual marker, participants drifted out of the cypher formation, unsure of where to stand or how to engage. We addressed this by physically marking the zones with tape, which restored some spatial legibility (Figures~\ref{fig:group2_noscreen} and \ref{fig:group3_noscreen}). Across the sessions, we also observed participants grooving more freely and expansively, suggesting that removing the screen invited fuller embodied attention even as it raised the initial social cost.

These observations suggested a design direction for the next iteration: replacing or augmenting the screen with projected light onto dancers themselves, with OSC-controlled position and intensity reflecting the same information the screen displays. Light projection would replace the screen's role as a shared focal point while keeping participants' attention on each other and the embodied space. However, this requires venue-specific lighting infrastructure that constrains where and how the system can be deployed. Given access to a venue with programmable stage lighting, we deployed this configuration in a public live performance (Section~\ref{sec:performance}).

\paragraph{Embodied Locus of Engagement}
Aggregated body-map responses (Figure~\ref{fig:bodymap}) reveal that participants most often felt change and engagement in their torso, hips, and feet---consistent with the rocking and stepping vocabulary of grooving that the system was designed to support. This embodied distribution is consistent with the design choice to use upper-body center-of-mass amplitude as the primary movement feature, though hips and feet lie outside what COMAmp measures.

\subsection{Results Summary}

Across both quantitative and qualitative data, three findings stand out. \textbf{Shared agency was the dominant experience}: strong agreement on Collaboration (81.8\%) and the collective-orchestration theme show that Encypher shifted agency from individual to group, even as individual Control scored more modestly (63.6\%)---a trade-off that supports the design intent. \textbf{Social presence was supported but not uniformly}: the screen gave participants a shared focal object that eased entry into the group but pulled attention away from one another; without it, embodied co-presence rose along with initial awkwardness. \textbf{Spatial markers shaped participation}: Spatial Awareness drew the highest ``strongly agree'' rate (36.4\%), and the marked spots acted as both clarity and constraint. Together, these results suggest that generative music can function as a social bridge for unacquainted participants, while surfacing a tension between visual cues and direct interpersonal attention.

\section{Public Museum Event} %
\label{sec:museum}

To test Encypher beyond a classroom setting and with members of the broader public, we deployed the system as part of \textit{We the Beat},
a 65-minute session within \textit{Design Redefined}, a public co-design
event hosted at the MIT Museum, centered around the theme of music, AI, and community.
The session was framed around the participatory design rule of \textit{Participatory Means} and was open to attendees who self-selected into the room from a set of three concurrent music workshops. Participants ranged from teens to adults with no expectation of dance experience. The session combined a brief introduction and demonstration with two longer phases of open exploration and collective performance design, ending with a closing showcase in the main event space.

\begin{figure}
    \centering
    \includegraphics[width=\linewidth]{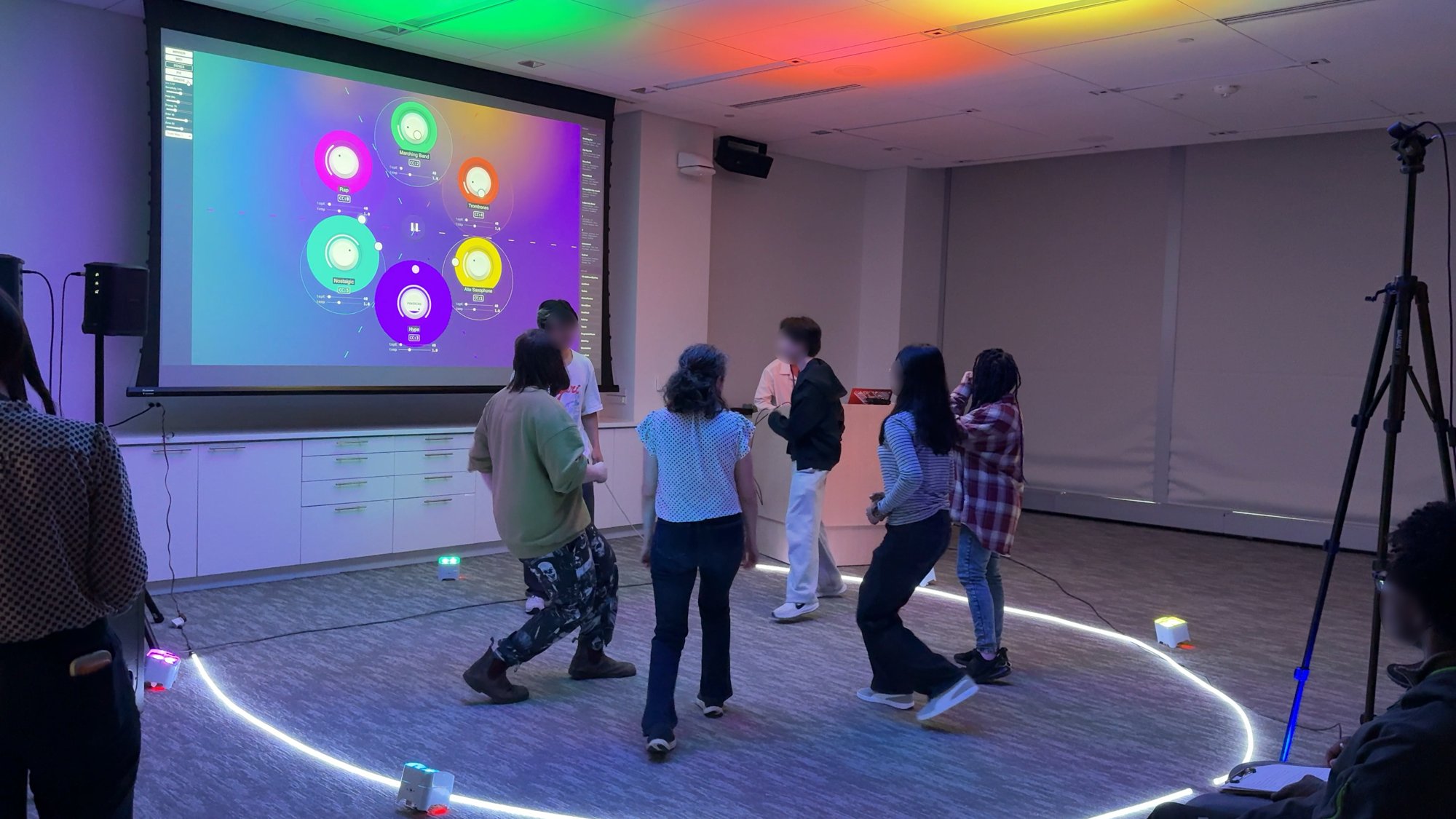}
    \caption{Strangers dancing with Encypher at the public museum event.}
    \Description{Six people dancing in a dark room with colorful lights projected onto the ceiling and a visualization on the big screen.}
    \label{fig:museum_teaser}
\end{figure}

\subsection{Setup}

The installation occupied an open circular space seven meters in diameter, configured for about 25 attendees in total, rotating through the cypher in groups of five to six (Figure~\ref{fig:museum_setup}, Figure~\ref{fig:museum_teaser}). The same depth-camera and LyriaRT pipeline used in the user study was deployed. For the event, we complemented the single large projection screen with mobile colored lighting projected on the ceiling, corresponding to the same colors as the on-screen visualization, rather than marked spots on the floor. We also dimmed the overall room lighting to evoke the atmosphere of an artistic or social dance space, closer to the conditions of an open cypher session than a classroom. Together, these changes served as a first deployment test of the ambient-light direction we had identified, in a venue with suitable lighting infrastructure.

\subsection{Observations}
Initially, participants were visibly hesitant to step into the cypher---an expected response to being asked to move with strangers in a public space. However, once one or two participants took the first step, others followed, and within a minute the room transitioned from observers to active dancers. The system appeared to function as the social bridge we had hypothesized, with the dance floor itself becoming the invitation rather than verbal prompting from facilitators.

We began the session with the screen on. As in the user study, participants kept looking at it, so partway through we disconnected the screen and ran the remainder with ceiling lighting alone. The light-based feedback proved sufficient: without the screen, participants still located their zones, moved freely within them, and coordinated with one another. The transition we observed in the classroom study---from screen-mediated attention to direct interpersonal awareness when the screen was removed---did not produce the same initial awkwardness here, suggesting that ambient, embodied feedback (light projected onto the dancers and the floor) can preserve the screen's coordinating function while keeping attention on the embodied space and other people. This deployment suggests that light projection is a viable replacement for screen-based visualization in venues where lighting infrastructure is available.

Two comments from attendees pointed toward future deployment contexts. One participant remarked, \textit{``I can totally see this in a club,''} positioning the system in nightlife and social-dance settings rather than only in research or pedagogical contexts. Another offered an analogy: \textit{``You know how there's karaoke machines and they're so popular---I imagine this to be similar.''} Both comments locate the value of the system not in the technical novelty of generative music but in its capacity to lower the social barrier to collective expression in informal, public settings. The karaoke comparison is particularly fitting: like karaoke, Encypher provides a shared, slightly silly, low-stakes structure that gives strangers permission to perform together. We did not design for nightlife or for the karaoke analogy explicitly, but both directions surfaced organically in a public deployment, suggesting that the system's social affordances may extend beyond the controlled study population.

The session culminated in a final debrief where attendees from all three concurrent workshops---approximately 80 people total---gathered in the main event space. Six participants from our session, \textit{We the Beat},
volunteered to demonstrate the system to the larger audience, performing a short collective piece that presented Encypher in front of a crowd that had not been part of the dance experience. The willingness of participants who had only encountered the system an hour earlier to perform publicly speaks to the ease with which Encypher welcomes newcomers into shared embodied performance, and suggests that the social presence and shared agency cultivated within the small-group session can carry into a broader performative context.

\section{Public Live Performance}
\label{sec:performance}

The screen's double role as social mediator (Section~\ref{sec:qualitative}) and the ceiling lighting at the museum event both pointed toward a version of Encypher with no screen at all, in which the visualization is projected onto the dancers themselves. We performed this version live at the MIT Music Technology Research Showcase in Tull Concert Hall on May 13, 2026,
in front of an audience of approximately 300 (Figure~\ref{fig:teaser}).

\subsection{Setup}

The performance followed a two-minute talk introducing the system. The performance used the same six stationary zones as the earlier deployments, each pre-assigned a musical element---dance jazz, Afrohouse, djembe, Hip-Hop, 808 drum machine, and classical piano---selected for how well they layered---a judgment informed by the co-design sessions and the museum event---and matched where possible to the performing dancers' own interests and training, so that each carried an element close to their own style. The screen, however, was removed entirely. Each zone was lit from above by a colored stage light, and the Max/MSP hub sent each dancer's COMAmp energy over OSC (via Ethernet) to the venue's lighting console (ETC Eos), which mapped it to the intensity of that dancer's light. A dancer charging their zone therefore brightened the pool of light they stood in, realizing the ambient-feedback direction that the museum event had only approximated. The depth camera was mounted 2.2\,m high, 3.5\,m from the center of the circle. Six dancers performed: five members of the local dance community and the first author, who joined for the final section. Preparation consisted of a single one-hour rehearsal devoted to blocking rather than choreography. The agreed structure was a staggered sequence of entries, beginning with Afrohouse and ending with dance jazz: one dancer freestyled alone, a second joined so that two musical elements blended, the first stepped out, and so on, building until every dancer was on stage and every element sounded together. Within that structure, all movement was improvised to the improvised music.

\subsection{Observations}

Unlike the museum event, this was a performance for a seated audience rather than a participatory session, and we did not collect participant or audience feedback; we report only the first author's observations. Without a screen there was no shared object to watch. Dancers oriented toward one another and toward the light at their feet, and the mutual attention that the screen had partly displaced during the user study was visibly restored. The pools of colored light also made each dancer's contribution legible to the audience in a way the screen never could, since energy showed up on the body itself. Because each element matched its dancer's training, the staggered entries read as a meeting of dance backgrounds as much as of musical genres, and in the final section, with all six dancers on stage and every element sounding, each dancer's style remained visible within the blended music rather than being flattened by it. Whether this arrangement changes how participants experience shared agency and social presence relative to the screen remains an open empirical question for a controlled study.

\section{Discussion}

\subsection{Designing With Versus Designing for a Community}

Our findings complicate the recently published C-Battle work \cite{grebel_designing_2026}, which reports largely positive engagement with movement sonification across novice, amateur, and one expert Hip-Hop dancer. We found something different: while novices and unacquainted participants embraced Encypher as a social bridge, established breaker communities resisted the system's framing. The discrepancy points to a design difference: the population a system is designed \textit{with}---and the population it is designed \textit{for}---substantially shapes what the system should be trying to support.

C-Battle's collective scenarios were evaluated with novice and amateur cohorts who had no reported pre-existing social ties; its single expert breakdancer tested only the individual scenarios. Its positive findings on validation and constraint-driven exploration are therefore findings about how individuals and small groups of dancers, mostly outside their home communities, engage with a sonification system.

Our co-design with breakers in their own studio sessions surfaced a different concern. Established communities have their own social dynamics and their own musical requirements: breakers need structured, predictable music with a clear downbeat to time freezes, headspins, and power moves. Beyond preference, there is a practical tension---over-sonifying movement risks turning the experience from dancing \textit{to} music into playing an instrument, subordinating the dance to the system rather than the other way around. For a community whose practice is already rich in social ritual, Encypher added overhead without adding value.

This points to a difference in goals. C-Battle is designed to extend and enrich an existing practice---giving trained Hip-Hop dancers a new expressive dimension within a vocabulary they already command. Encypher's goal is different: to introduce people to the joy of dancing and to help strangers bond through shared movement. It has evolved to be less a tool for dancers and more a social invitation extended through music---a low-stakes entry point that lowers the threshold for moving together in public. The pivot from breakers to unacquainted participants was therefore not a retreat but a recognition of what the system is actually for.

\subsection{Mapping as Social Design}

One of the clearest lessons from building Encypher across multiple iterations is that the choice of mapping is not a technical detail---it is a social one. Every mapping encodes an implicit theory of how people should interact: who is heard, what behaviors are rewarded, and what collective behavior improves the music. Designing a mapping where scattered individuals produce chaotic sound and a tight circle produces coherence is, in effect, designing a social norm into the system. The music becomes a mechanism for shaping group behavior---a form of social choreography through sound.

A social framing suggests mappings that Encypher did not yet explore. The question of what happens when multiple participants converge on the same zone---whether to blend their prompts, amplify the combined energy, or introduce tension---was raised by participants during co-design but never implemented. Future work should treat spatial coincidence and convergence, alongside synchrony (Section~\ref{sec:future}), as leading design parameters, each affording distinct forms of group experience.

More broadly, we see the mapping design process as iterative and deeply embodied. What constitutes ``energy'' in movement---the quality that draws people together and compels a crowd to watch---varies substantially across dance styles, skill levels, and cultural contexts. COMAmp captures one facet of groove, but it is not the full story. Movement energy is still poorly understood as a mappable signal, and discovering what connects the feel of a moment to the music that fits it has required slow, collaborative experimentation with real dancers across many sessions. That slowness is not a limitation; it is part of the process.

\subsection{Symbolic Versus Neural Audio as Design Modalities}

Encypher and \textit{Con Moto} \cite{chen_con_2026} represent two distinct modalities for generative music in dance contexts, and comparing them shows how the underlying model changes what can be designed.

\textit{Con Moto} operates on symbolic MIDI generation. This gives designers fine-grained control: individual voices can be steered, pitch ranges masked, harmonies conditioned, and spatial audio positioning applied per instrument across an eight-channel ambisonic field. The result is a system capable of tight performance choreography---every musical dimension can be tuned, and the two-person arc of the piece was designed around that expressive resolution. For performances with known participants, defined roles, and a rehearsal process, symbolic generation offers a high degree of craft.

Encypher operates on neural audio. The musical output is substantially richer in texture and danceability---LyriaRT's training on large-scale studio recordings produces music that is more compelling for movement, unconstrained by the note-throughput limits that cap MIDI generation's density and diversity. Text-based prompt steering also opens a different kind of creative space: participants can articulate what they want the music to sound like in natural language, which turned out to be a generative design move in itself---asking dancers what genres and instruments they wanted surfaced preferences and opened collaborative conversations that note-level control never would. The trade-off is that neural audio loses spatialization, fine-grained note-level control, and some of the predictability that makes performance choreography tractable.

We see these not as competing approaches but as occupying different positions between controllability and richness. Symbolic generation excels at controllability; neural audio excels at richness. Which matters more depends on whether the system is designed for performance (where control enables craft) or for social participation (where richness enables pleasure and danceability). Encypher's finding that novices and strangers responded strongly to the quality of the music---not just to having agency over it---suggests that for social participation contexts, richness may matter more than control, and that trading fine-grained mappings for high-fidelity audio is a defensible constraint.

\section{Limitations and Future Work}
\label{sec:future}

\paragraph{Study Limitations}
The user study involved eleven participants in a single session, evaluated through self-report and the first author's observations, with the first author present as a guest instructor. The screen and no-screen conditions were not counterbalanced: removing the screen coincided with group order and with the introduction of floor tape, so the shift we attribute to the screen cannot be cleanly separated from those factors. The museum event and live performance add ecological range but were observed rather than measured. The findings should therefore be read as design evidence from a first deployment cycle rather than as controlled effects. A counterbalanced screen/no-screen comparison, and feedback from participants and audience in the performance setting, would address these gaps.

\paragraph{Baseline, Model, and Transfer}
There was also no baseline condition. Our original study design compared cyphers under a pre-selected playlist, a live DJ, and Encypher, but we set it aside: a comparison between a first prototype and a long-standing practice would have been premature, and building the system with dancers first was the more informative use of the sessions. With the system now stable, that comparison is the appropriate next study, and until it is run the shared agency participants reported cannot be attributed to generative music specifically rather than to shared music in a cypher. LyriaRT is a closed model accessed through an API whose behavior may change, so the musical output is not exactly reproducible; the open-weights MagentaRT offers a reproducible alternative, though we did not evaluate it. Finally, the mappings are community-specific by design: \textit{Con Moto}'s limb-velocity mappings, designed for a duet performance over classical piano and jazz models, did not survive co-design with breakers and were replaced by COMAmp, and we expect other dance forms to require the same re-design rather than a direct transfer.

\paragraph{Latency and Rhythmic Grounding}
LyriaRT's two-second generation window creates a perceptible delay between movement and audio response that disrupts the causal coupling dancers expect. Compounding this, LyriaRT does not maintain consistent meter and downbeat alignment---a serious limitation for breaking dancers, for whom hitting the downbeat is fundamental to technique. Text-based BPM steering proved unreliable: genre and instrument combinations often refused to lock to a target tempo, making rhythmic predictability difficult to guarantee. The most promising model-level fix would be a beat-conditioning input injected at inference time---similar to audio injection already supported in MagentaRT---that allows the system to synchronize generation to a detected or user-specified pulse. As LyriaRT and similar real-time audio models mature, we suggest that beat conditioning, not just text prompts, be treated as a control signal for dance contexts.

\paragraph{Synchrony as an Unmapped Social Dimension}
Synchrony between participants---moving in phase with one another---is a well-established driver of social bonding \cite{tarr_synchrony_2015, rabinowitch_musical_2021}, yet it was absent from Encypher's mapping design. Defining synchrony as a computable signal is non-trivial: it requires choosing a timescale (beat-level, phrase-level, or postural drift), a body region (limb velocity, torso sway, or foot timing), and a comparison metric (cross-correlation, phase coherence, or mutual information). Different definitions would reward different social behaviors---brief rhythmic alignment vs.\ sustained ensemble coherence---and each would afford a distinct group experience. Future work should treat synchrony as a primary mapping parameter, designing conditions where two or more dancers moving in sync produces a richer or more stable musical output, building on the idea that musical output is an incentive for social presence.

\paragraph{Ambient Feedback and Lighting as a Control Surface}
The museum deployment demonstrated that colored ceiling lighting, driven by the same OSC signals as the screen visualization, could preserve spatial and energetic legibility without requiring participants to attend to a separate display, and the live performance (Section~\ref{sec:performance}) removed the screen entirely, driving the intensity of each dancer's light from their own energy. Automated lighting creates a multi-sensory feedback loop that reinforces the embodied experience without pulling attention away from other dancers. What remains untested is the group level---collective energy building toward a peak that floods the room with color---and a controlled comparison of screen-based and light-based feedback on social presence. Implementation requires DMX or networked lighting control integrated into the Max/MSP hub, which is feasible in most performance venues but constrains deployment to spaces with programmable stage lighting.

\paragraph{Toward a Stage That Listens to Itself}
In the live performance, the venue itself became part of the feedback loop: dancers moved, the camera sensed, the music and the stage lights responded, and the dancers moved differently in turn. Nothing in this loop was specific to Encypher's six zones, nor to LyriaRT: the sensing pipeline, social-geometry metrics, and mapping principles carry over to any real-time generative backend. A stage instrumented with sensors, whose lighting, sound, and generative systems each act autonomously on shared movement data, generalizes the idea from a single installation to a design pattern for performance spaces. Future work could deploy such a sensing stage at larger scale---more performers, more output modalities, longer runs---and study how a room that responds to its occupants reshapes the social behavior within it.

\section{Conclusion}

We presented Encypher, a collaborative generative music system that translates collective movement qualities into text prompts to condition real-time neural audio generation for dance cyphers. Built through autobiographical design, five weeks of co-design with local breakers, an IRB-approved user study with eleven unacquainted participants, a public museum deployment, and a public live performance, Encypher reframes generative music in HCI from a tool for the solo performer to a medium that can shape social experiences in shared physical space.

Across these settings, three findings stand out. Mapping group-level features---collective energy and spatial circularity---rather than individual gestures shifted the felt locus of agency from the self to the room. The system functioned as a social bridge for strangers, while established dance communities, with their own social dynamics and stricter musical requirements, sometimes resisted its framing. And the visualization screen acted as a double-edged mediator, easing entry into the group at the cost of direct interpersonal awareness---a tension that ambient lighting at the museum and the screen-free live performance progressively resolved.

These findings argue for treating mapping design as social design: every choice about what the system listens for, what it rewards, and what it ignores encodes an implicit theory of how people should interact. The contributions of this work---the Encypher system, empirical evidence of shared agency emerging from collective movement features, and design implications for collaborative human-AI creative systems---offer a starting point for AI systems that support not only what we can compose, but how we come together to dance.

\begin{acks}
The work was supported by the MIT MGAIC Research Fund and the Quanta Computer Graduate Fellowship. The authors would like to thank the members of the MIT Imobilare and moveMENtality dance groups for their collaboration during weekly co-design sessions and for their insightful feedback. We are grateful to Grisha Coleman for providing expert perspectives as a dance and technology practitioner and for facilitating a study within her classroom. We also thank the MIT Museum, Morningside Academy for Design, and Innovators for Purpose for organizing the Design Redefined event and coordinating the complex logistics involved. We thank Eran Egozy for feedback on the system and the live performance, and for organizing the Music Technology Research Showcase with the MIT Music and Theater Arts faculty. We thank the dancers who performed in the showcase, Perry Naseck for helping with the lighting, and Alex Tung and Mark Rau for helping with the sound. Finally, we thank Kimberly Y. Zhang for her invaluable assistance during the user study and public museum event, as well as for her participation as a system user.
\end{acks}

\bibliographystyle{ACM-Reference-Format}
\bibliography{references}

\appendix

\section{Post-Session Questionnaire}
\label{app:questionnaire}

The following questionnaire was administered on paper to participants immediately after each Encypher trial during the user study. The survey was anonymous, identifying participants only by Participant ID and group size. Participants also indicated whether they had danced in a group setting before.

\subsection{Section 1: Experience of Dancing}

Participants rated each statement on a 7-point Likert scale from 1 (Strongly Disagree) to 7 (Strongly Agree).

\begin{enumerate}
    \item \textbf{Engagement:} I found dancing with this system fun and engaging.
    \item \textbf{Control:} I felt that I had agency over the musical output.
    \item \textbf{Enjoyment:} The musical output was pleasing and I enjoyed hearing it.
    \item \textbf{Replay:} I would like to try Encypher again.
\end{enumerate}

\subsection{Section 2: Social Interaction}

Participants rated each statement on a 7-point Likert scale from 1 (Strongly Disagree) to 7 (Strongly Agree).

\begin{enumerate}
    \setcounter{enumi}{4}
    \item \textbf{Social Presence:} The system made me more aware of the other people in the space and what they were doing.
    \item \textbf{Collaboration:} I felt we were sharing energy or a collective vibe as a group and that we were collaborating to create the music.
    \item \textbf{Invitation:} I felt prompted or encouraged by the music or another dancer to participate.
    \item \textbf{Encouragement:} I felt motivated to keep dancing because of others'/the system's response.
    \item \textbf{Spatial Awareness:} The marked spots in the space made it clear where I could go and what role I could take.
\end{enumerate}

\subsection{Section 3: Embodied Reflection}

Open-ended prompts:

\begin{enumerate}
    \setcounter{enumi}{9}
    \item \textbf{Body Map:} On the outline provided, circle or shade the part(s) of your body where you felt most engaged or noticed something change during the session. Add a word or phrase next to any area.
    \item \textbf{Connection:} Was there a moment where you felt connected to another dancer or to the music? Describe what happened physically or sonically, and what nonverbal cues did you use?
    \item \textbf{The Spots:} Did the marked spots in the space affect how you moved or how you interacted with others? If so, how?
    \item \textbf{Agency:} Did you feel like you were playing the music, or did the music feel like it was responding to something larger---the group, the room? Describe the feeling.
    \item \textbf{Hesitation:} Was there a moment you wanted to join in or take a spot but held back? What stopped you?
    \item \textbf{Anything else:} Anything you noticed, felt, or want to tell us that the questions above didn't capture.
\end{enumerate}

\subsection{Group Debrief Guide}

Following individual surveys, a 20-minute facilitated group discussion was audio-recorded with consent. Discussion prompts were directed to the group rather than individuals, encouraging accounts to build on one another:

\begin{itemize}
    \item Was there a moment where you felt connected to someone else in the space? What happened?
    \item Did you feel like the music belonged to you, to the group, or to something else?
    \item Was there a moment you wanted to join in but didn't? What held you back?
    \item Did the spots change how you moved or where you paid attention?
    \item Did the system make the space feel more or less welcoming to joining in?
\end{itemize}

Researchers tracked observation codes in real time, including spot occupancy, eye contact, approach/retreat behavior, laughter and smiling, and timestamps of musical shifts.

\end{document}